\documentclass[usenatbib]{mnras}
\usepackage[english]{babel}
\usepackage[utf8]{inputenc}
\usepackage[T1]{fontenc}
\usepackage{graphicx}
\usepackage{amsmath}
\usepackage{amssymb}
\usepackage{hyperref}
\usepackage{etoolbox}
\makeatletter
\patchcmd\@combinedblfloats{\box\@outputbox}{\unvbox\@outputbox}{}{%
\errmessage{\noexpand\@combinedblfloats could not be patched}%
}%
\makeatother
\usepackage{wasysym}
\usepackage{threeparttable}
\usepackage{pdflscape}
\title[An eruptive phase heralded the type IId SN 2013gc]{Signatures of an eruptive phase before the explosion of the peculiar core-collapse SN 2013gc}
\author[Reguitti et al.]{
Andrea Reguitti$^{1}$\thanks{E-mail: andrea.reguitti@studenti.unipd.it},
A. Pastorello$^{2}$, G. Pignata$^{3,4}$, S. Benetti$^{2}$, E. Cappellaro$^{2}$, \newauthor
\ M. Turatto$^{2}$, C. Agliozzo$^{3,4}$, F. Bufano$^{5}$, N.I. Morrell$^{6}$, F. Olivares E.$^{3,4}$, \newauthor
\ D.E. Reichart$^{7}$, J.B. Haislip$^{7}$, V. Kouprianov$^{7}$, S.J. Smartt$^{8}$, \& S. Ciroi$^{1}$\\
$^{1}$Dipartimento di Fisica e Astronomia 'G. Galilei', Università di Padova, Vicolo dell'Osservatorio 3, 35122 Padova, Italy\\
$^{2}$INAF – Osservatorio Astronomico di Padova, Vicolo dell'Osservatorio 5, 35122 Padova, Italy\\
$^{3}$Departamento de Ciencias F\'{i}sicas – Universidad Andrés Bello, Avda. Rep\'{u}blica 252, Santiago, Chile\\
$^{4}$Millennium Institute of Astrophysics, Nuncio Monsenor S\'{o}tero Sanz 100, Providencia, Santiago, Chile\\
$^{5}$INAF – Osservatorio Astrofisico di Catania, Via S. Sofia 78, 95123 Catania, Italy\\
$^{6}$Las Campanas Observatory, Carnegie Observatories, Casilla 601, La Serena, Chile\\
$^{7}$Department of Physics and Astronomy, University of North Carolina at Chapel Hill, Campus Box 3255, Chapel Hill, NC 27599, USA\\
$^{8}$Astrophysics Research Centre, School of Mathematics and Physics, Queen’s University Belfast, Belfast, BT7 1NN, UK\\}
\date{Accepted 2018 October 21. Received 2018 September 27; in original form 2018 June 21}
\begin{document}
\maketitle
\volume{00}
\pagerange{1-20}
\pubyear{2018}

\begin{abstract}
We present photometric and spectroscopic analysis of the peculiar core-collapse SN 2013gc, spanning seven years of observations. The light curve shows an early maximum followed by a fast decline and a phase of almost constant luminosity. At +200 days from maximum, a brightening of 1 mag is observed in all bands, followed by a steep linear luminosity decline after +300 d. In archival images taken between 1.5 and 2.5 years before the explosion, a weak source is visible at the supernova location, with mag$\approx$20.
The early supernova spectra show Balmer lines, with a narrow ($\sim$560 km s$^{-1}$) P-Cygni absorption superimposed on a broad ($\sim$3400 km s$^{-1}$) component, typical of type IIn events. Through a comparison of colour curves, absolute light curves and spectra of SN 2013gc with a sample of supernovae IIn, we conclude that SN 2013gc is a member of the so-called type IId subgroup.
The complex profile of the H$\alpha$ line suggests a composite circumstellar medium geometry, with a combination of lower velocity, spherically symmetric gas and a more rapidly expanding bilobed feature. This circumstellar medium distribution has been likely formed through major mass-loss events, that we directly observed from 3 years before the explosio. The modest luminosity ($M_I\sim-16.5$ near maximum) of SN 2013gc at all phases, the very small amount of ejected $^{56}$Ni (of the order of $10^{-3}$ M$_\odot$), the major pre-supernova stellar activity and the lack of prominent [\ion{O}{I}] lines in late-time spectra support a fall-back core-collapse scenario for the massive progenitor of SN~2013gc.
\end{abstract}

\begin{keywords}
supernovae: general, supernovae: individual: SN 2013gc, SN 1994aj, SN 1996al, SN 1996L, SN 2000P
\end{keywords}

\section{Introduction}
\label{introduction}
Major astronomical surveys, such as the Asteroid Terrestrial-impact Last Alert System \citep[ATLAS;][]{2018arXiv180200879T}, the All Sky Automated Survey for SuperNovae \citep[ASAS-SN;][]{2014AAS...22323603S} and the Panoramic Survey Telescope and Rapid Response System \citep[Pan-STARRS;][]{2016arXiv161205560C} are discovering a growing number of peculiar stellar transients that display a wide range of photometric and spectroscopic properties. A fraction of them show supernova (SN) features, with high velocity gas. Some exhibit also signatures of interaction between fast-moving ejecta and circumstellar medium (CSM), which is revealed through multi-component line profiles in the spectra. In particular, supernovae (SNe) showing narrow H emission lines likely produced in the unshocked photoionised gas \citep{1994ApJ...420..268C,1994MNRAS.268..173C} are classified as type IIn \citep{schlegel}. A well-studied member of this class is SN 1988Z \citep{1993MNRAS.262..128T,1991MNRAS.250..786S,1999MNRAS.309..343A,2017MNRAS.466.3021S}.

Dense H-rich CSM is generally produced through mass-loss events from the progenitor star, which occurred from tens to thousands years before the SN explosion \citep{Smith2014}. However, in some cases, H-rich CSM grows through eruptive episodes occurred a short time (e.g., a few months to years) before the terminal stellar death. The energetics of these pre-SN outbursts is quite high ($10^{48}-10^{49}$ erg), and the ejected gas  moves at a speed of tens to $\lesssim$ 10$^3$ km s$^{-1}$. In contrast, fast-moving SN ejecta have velocities of several thousands km s$^{-1}$.

The photometric evolution of SNe IIn is different from that of normal type II SNe. The recombination of the CSM photoionized by the shock breakout gives an additional source of energy, powering the very early light curve of the SN. The subsequent interaction between SN ejecta and CSM converts the kinetic energy of the shock wave into radiation. This is a more efficient powering mechanism, allowing the luminosity to remain almost constant. In some cases, the light curve may even show a brightening. This source of energy is additive to the $^{56}$Ni/$^{56}$Co decays, and the SN remains visible even for years after the explosion \citep{Milisavljevic2012,2016MNRAS.456.3296B}.

The variety of observational properties of SNe IIn depends on the velocity of the different gas components, their chemical composition and geometry, the mass-loss history and consequently the CSM density profile \citep{1997ARep...41..672C}.
Their spectra are characterized by strong emission lines of the Balmer series superposed on a rather blue continuum, with H$\alpha$ being the most prominent feature. The profile of these lines is complex, including a broad shallow component, on top of which the characteristic narrow line sits. The velocities inferred from the full-width-at-half-maximum (FWHM) of the 2 components are typically several thousands (sometimes exceeding $10^4$), and few hundreds km s$^{-1}$, respectively. In some cases a blue-shifted narrow P-Cygni absorption is also present. A third component, with intermediate width, can also be observed with a velocity of several hundreds to a few thousand km s$^{-1}$. The H$\alpha$ line dominates the spectra at all phases, including at late times, when the blue continuum becomes redder or eventually disappears. 

Different types of massive core-collapse SN progenitors are known to experience severe mass loss during their late evolutionary phases, including events triggered by major outbursts. Luminous and massive stars, including hypergiants and red supergiants, may lose mass through strong stellar winds during their evolution, or via binary interaction.
In particular, luminous blue variables (LBVs) occasionally produce a spectacular giant eruption (GE), during which they lose a considerable amount of material (up to tens of solar masses). The best followed GE of an LBV in the Milky Way was that of $\eta$ Carinae in the mid 19th century. It lasted a few decades, during which the star grew up its luminosity up to absolute magnitude $M\sim-14$ \citep{2011MNRAS.415.2009S}. An LBV GE which occurs in a distant galaxy can be confused with an under-luminous type IIn SN. For this reason, extra-galactic GEs (or major outbursts) of hypergiants are sometimes dubbed as "SN Impostors" \citep{2000PASP..112.1532V}.

LBVs may exhibit an extreme variability, with oscillations exceeding 5 mag, over periods of many years. These outbursts sometimes are premonitions of subsequent type II SN explosions. This was directly observed in SN 2009ip \citep{2013MNRAS.430.1801M,2014MNRAS.438.1191S}; LSQ13zm \citep{2016MNRAS.459.1039T}, SN 2016bdu \citep{2018MNRAS.474..197P}; SNhunt151 \citep{2018MNRAS.475.2614E}. The progenitor of SN 2009ip had a non-terminal outburst in 2009. Other outbursts of the same object were observed in the following years, until a much brighter event occurred in summer 2012, which was claimed to be the signature of the final core-collapse. Whether the 2012 event was a real SN or a non-terminal explosion is still controversial \citep{2013ApJ...767....1P,2014ApJ...780...21M}.
The sequence of events observed in SN 2009ip indicated that the progenitor was likely an LBV. Multiple outbursts accompanied by major mass loss allow the formation of a dense H-rich cocoon around the star. Hence, when the star finally exploded, it showed up as a type IIn SN.

In the vast array of displays of SNe IIn, a group exhibits peculiar "double" (broad and narrow) P-Cygni profile in the H lines, signatures of SN ejecta and CSM, respectively. 
For this spectral property, these peculiar type IIn events are labelled as "type IId" \citep[see the review of][]{2000MmSAI..71..323B}. This category includes SN 1994aj \citep{1998MNRAS.294..448B}, SN 1996L \citep{1999MNRAS.305..811B} and SN 1996al \citep{2016MNRAS.456.3296B}. These objects are quite rare, and this paper discuss the case of a possible new member, SN 2013gc.

The structure of the paper is as follows: we introduce the discovery of SN 2013gc, the physical properties of the host galaxy, including a discussion on the reddening and the distance, in Sect. \ref{discover}.
Information of the instrumentation used and a description on the data reduction techniques are reported in Sect. \ref{setup}. An analysis of the light curve is provided in Sect. \ref{light}, while in Sect. \ref{colour} we analyse the reddening-corrected colour and absolute magnitude curves. In Sect. \ref{historia} we present robust evidence of progenitor variability before the explosion. The spectral features and evolution of H$\alpha$ profile are discussed in Sect. \ref{spectra}. A discussion on the progenitor and the physics of the explosion is presented in Sect. \ref{discussion}. Finally, the conclusions are summarized in Sect. \ref{conclusion}.
In appendix \ref{2000p}, we present the photometric measurements and the spectra of another SN IIn/IId 2000P, which has been used as a comparison object in this paper.

\begin{table}
\label{tab1}
\caption{Properties of ESO 430-20}
\begin{threeparttable}[b]
\begin{tabular}{ll}
\hline
$\alpha$ (J2000)    & 08$^h$07$^m$08$^s$.6 \\
$\delta$ (J2000)   	& -28$^{\circ}$03'08".8 \\
morphological type	& SAB(s)d \\
Diameter\tnote{1}	& 1.90$\times$0.78 arcmin \\
redshift\tnote{1}	& 0.003402$\pm0.000007$ \\
v$_{Hel}$\tnote{1}	& 1020$\pm2$ km s$^{-1}$ \\
v$_{Virgo+GA+Shapley}$\tnote{2}	& 942$\pm20$ km s$^{-1}$ \\
Distance Modulus\tnote{3}	& 30.46$\pm0.19$ \\ 
$A_V$				& 1.253 mag \\
\hline
\end{tabular}
\begin{tablenotes}
\item[1] \citet{1998AAS..130..333T} \item[2] \citet{2000ApJ...529..786M} \item[3] adopted
\end{tablenotes}
\end{threeparttable}
\end{table}

\section{Discovery}
\label{discover}
\begin{figure}
\includegraphics[width=\columnwidth]{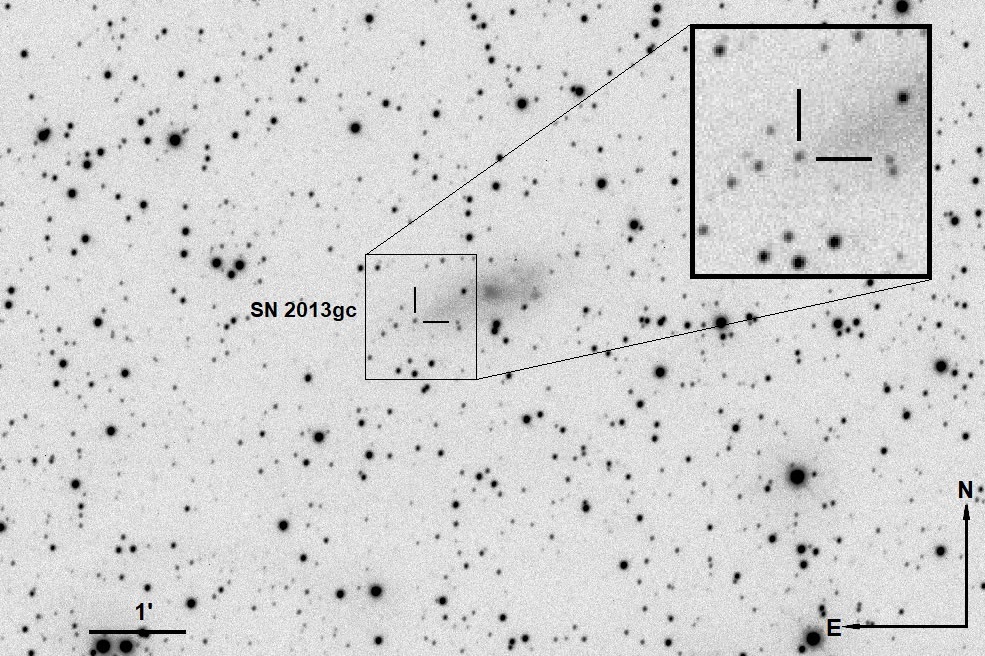}
\caption{Discovery image of SN 2013gc taken on 2013 November 7 with the PROMPT-3 telescope and clear filter. A blow up of the SN location is shown in the top-right inset.}
\label{discovery}
\end{figure}

SN 2013gc (=PSN J08071188-2803263) was discovered on 2013 November 7 at RA=$08^h$07$^m$11$^s$.88 and Dec=$-28^{\circ}$03'26".32 (J2000), 43.3" east and 18.0" south of the nucleus of the galaxy ESO 430-20 \citep[a.k.a. PGC 22788,][]{2013CBET.3699....1A}. The discovery chart is in Fig. \ref{discovery}.
The discovery was done by the CHilean Automatic Supernova sEarch survey \citep[CHASE;][]{2009RMxAC..35R.317P}, using the Panchromatic Robotic Optical Monitoring and Polarimetry Telescopes \citep[PROMPT;][]{2005NCimC..28..767R}, at the Cerro Tololo Inter-American Observatory (CTIO). 

The original spectral classification indicated it was SN IIn \citep{2013CBET.3699....1A}. The spectrum, discussed in Sect. \ref{spectra}, was obtained soon after the discovery with the Las Campanas 2.5-m du Pont telescope (+WFCCD), and cross-correlated with a library of SN spectra using the "Supernova Identification'' code  \citep[SNID;][]{2007AIPC..924..312B}. SN 2013gc appeared as a type IIn SN similar to SN 1996L \citep{1999MNRAS.305..811B} at about two months after the explosion.

\subsection{The SN environment}
The host galaxy ESO 430-20 is a member of a small group of galaxies \citep{2007ApJ...655..790C}.
The morphological classification, from the RC3 Catalogue \citep{1991RC3.9.C...0000d}, is SAB(s)d.
Details on the host galaxy obtained from the NASA/IPAC Extragalactic Database (NED)\footnote{http://ned.ipac.caltech.edu/} are summarized in Tab. \ref{tab1}.
The type II SN 2013ak was also discovered in this galaxy, on 2013 March 9 \citep{2013CBET.3437....1C}, at a magnitude of 13.5.
The location of SN 2013gc is far from the galaxy center: with a distance of 12.37 Mpc, the SN is 2.81 kpc from the nucleus, while the radius of the galaxy is 3.4 kpc. Due to the relatively large radial distance from the galaxy center, the local dust contamination is likely small.

\subsection{Distance}
The distance of ESO 430-20 is controversial.
The Virgo + Great Attractor + Shapley velocity ($v=942\pm20$ km s$^{-1}$, Tab. \ref{tab1}), assuming $H_0=73$ km s$^{-1}$Mpc$^{-1}$, provides a distance $d=12.90\pm0.27$ Mpc, hence $\mu=30.55\pm0.05$ mag.
\citet{2007ApJ...655..790C} determined a distance of 13.36 Mpc, equal to $\mu=30.63$ mag.
The HyperLeda catalogue reports a radial velocity corrected for Virgo Cluster infall $V_{Vir}=793\pm2$ km s$^{-1}$, hence a distance $d=12.0^{+10.6}_{-5.6}$ Mpc and $\mu=30.40\pm1.37$ mag.

\citet{2007A&A...465...71T} gave three different distance estimates of the host galaxy through photometry in 3 NIR bands, and applying the Tully-Fisher (TF) relation: $d=11.4$ Mpc in \textit{J}, $d=11.5$ Mpc in \textit{H}, $d=12$ Mpc in \textit{K}, with $\mu$ of 30.29$\pm0.46$, 30.31$\pm0.47$, 30.40$\pm0.45$ mag, respectively.

More recently, \citet{2016AJ....152...50T} estimated $d=6.95$ Mpc for ESO 430-20, corresponding to $\mu=29.21\pm0.54$ mag, obtained from the TF relation, calibrated on a sample of galaxy clusters. The discrepancy with the previous determinations is very large, so we do not take it in account this result in the choice of the distance to adopt.

For the distance, the mean of the three TF determinations obtained by \citet{2007A&A...465...71T}, and the two kinematical velocities reported in Tab. \ref{tab1} are considered.
We hence adopt $d=12.37\pm1.07$ Mpc and $\mu=30.46\pm0.19$ mag through out this paper.

\subsection{Reddening}
The Galactic reddening $A_V$ reported in Tab. \ref{tab1} is from \citet{2011ApJ...737..103S}. Assuming $R_V=3.1$ \citep{1999PASP..111...63F}, a colour excess $E(B-V)=0.404$ mag is derived. We adopt the \citet{1989ApJ...345..245C} extinction law. 
The extinction is quite large due to the low Galactic latitude of the galaxy ($\sim2.3^{\circ}$). This makes the reddening value very uncertain. We warn that the inaccuracy of the Schlafly extinction maps can lead to inaccuracy on the intrinsic distance and absolute magnitude of the object.

On the other hand, we are not able to estimate the internal reddening of the host galaxy. We notice that the spectra do not show the narrow \ion{Na}{I}D doublet features at the redshift of the host galaxy, because of the modest flux of the continuum at these wavelengths. The host galaxy reddening is hence assumed to be negligible, which is consistent with the remote location of the SN in the host galaxy.

\section{Setup and data reduction}
\label{setup}
Johnson-Cousins \textit{BVRIJHK}, Sloan \textit{grizy} and unfiltered photometric images were obtained with a large number of observing facilities, whose technical notes are here summarized.
For the photometry, we used:
\begin{itemize}
\item The PROMPT facility, located at CTIO, which consists of six 0.41-m robotic telescopes \citep{2005NCimC..28..767R}\footnote{The optical imagers, made by Apogee, use back-illuminated E2V 1024$\times$1024 CCDs, with a 10’ field of view and a pixel scale of 0.6”/pixel.}.

\item The TRAnsiting Planets and PlanetesImals Small Telescope (TRAPPIST) is a robotic 0.5-m Ritchey-Chrétien telescope located at ESO La Silla Observatory\footnote{The telescope is equipped with a Fairchild 3041 back-illuminated 2k$\times$2k CCD. The pixel scale is 0.64”/pixel, the field of view is 22’$\times$22’.}.

\item The NTT telescope at La Silla, equipped with the ESO Faint Object Spectrograph and Camera (EFOSC2)\footnote{The detector of the instrument is a thin CCD with 2048$\times$2048 pixels and a pixel size of 0.12".} and the Son of Isaac (SOFI)\footnote{This is a NIR spectrograph and imaging camera, with a HgCdTe 1024$\times$1024 pixels detector.}.

\item The Southeastern Association for Research in Astronomy (SARA) is a consortium of colleges that operates some remote facilities, including a 0.6-m telescope\footnote{The instrument is equipped with a 2048$\times$2048 pixel E2V CCD. The pixel scale is 0.44”/pixel, with a FOV of 15'. The filter set includes Johnson \textit{UBVRI} and \textit{SDSS ugriz}.} on Cerro Tololo, in Chile.

\item The SMARTS 1.3-m CTIO telescope, with the ANDICAM detector\footnote{Two instruments can perform optical and NIR images. The CCD is 1024$\times$1024 pixels wide, and has a pixel scale of 0.371”/pixel.}.

\item The Gemini-South 8.1-m telescope at Cerro Pach\'{o}n, with the GMOS instrument\footnote{The instrument is an imager, long-slit and multi-slit spectrograph, with a FOV of over 5.5 square arcminutes. It is equipped with a EEV CCD, with a pixel scale of 0.08"/pixel.}.

\item The 1.2-m "S. Oschin" Schmidt telescope at Palomar Observatory, during the Palomar Transient Factory (PTF) survey\footnote{The survey is performed with a 12k$\times$8k CCD mosaic. The pixel scale is 0.1"/pixel.}. These data were retrieved through the web interface \url{http://irsa.ipac.caltech.edu/applications/ptf/}.

\item The 1.8-m telescope on Haleakala, Hawaii, during the Pan-STARRS 1 (PS1) survey \citep{2016arXiv161205560C,1612.05240}, with the Gigapixel Camera 1 (GPC1)\footnote{The camera is composed of an array of 60 back-illuminated CCDs, with a 7 square degrees field-of-view. The pixel scale is 0.258"/pixel. The survey is performed with $grizy$ filters.}.

\item The European Southern Observatory (ESO) 2.6-m Very Large Telescope (VLT) Survey Telescope (VST) at Cerro Paranal, with OMEGACAM. The camera is composed with a mosaic of 32 2k$\times$4k pixels CCDs, with a total of 268 megapixels and a field of view of 1 square degree. The pixel scale is 0.215"/pixel.
\end{itemize}

\subsection{Data reduction}
For the photometric data reduction we used a dedicated pipeline called SNOoPY\footnote{SNOoPy is a package for SN photometry using PSF fitting and/or template subtraction developed by E. Cappellaro. A package description can be found at http://sngroup.oapd.inaf.it/snoopy.html.} \citep{cappellaro}.

The optical images were corrected for bias, overscan and flat-field, using standard IRAF tasks. If several dithered exposures with the same instrumental configuration were taken on the same night, they were combined  to increase the signal-to-noise ratio (SNR). The SNOoPy package is used for astrometric calibration and seeing determination on the images. 
The PSF-fitting technique was used to determine the instrumental SN magnitude. For each image, we built a PSF model using the profiles of bright, isolated stars in the field. We subtracted the sky background fitting a low-order polynomial (typically a second-order). Then, the modelled source was subtracted from the original frames, a new estimate of the local background is performed and the fitting procedure is repeated.
If the source is not detected, an upper limit to the luminosity is established.

Photometric nights were used to calibrate the magnitudes of reference stars in the SN field, through the observations of \citet{1992AJ....104..340L} standard fields with the same instrumental setup. This local sequence was used to correct the instrumental zero points in non photometric nights. Photometric errors were estimated through artificial star experiments, combined in quadrature with the uncertainties derived from the PSF-fitting returned by DAOPHOT.
Finally, we calibrated the final instrumental magnitudes of the SN in the Johnson-Cousins and Sloan photometric systems, using differential photometry. For the calibration of the PS1 survey images, we followed the prescriptions of \citet{1612.05242}.

The few Sloan $r$ and $i$ (apart those from PS1 survey) magnitudes were transformed in the Johnson-Cousins photometric system following the equations of \citet{2008AJ....135..264C}.
Clear filter magnitudes from PROMPT were treated as Cousins $R$-band magnitudes, because the wavelength efficiency peak of the detector is similar to that of the $R$ filter response curve.

Only a single epoch of $J, H$ and $K$ imaging of the SN, along with a few additional NIR frames of the SN field taken before the explosion, are available. Clean sky images were obtained by median-combining multiple dithered images of the field, and consequently subtracted to individual images. Sky-subtracted images were finally combined to increase the SNR. PSF-fitting photometry was also performed on the NIR sources, and the final SN magnitudes were calibrated using the 2MASS catalog as reference.

\section{Light curve}
\label{light}
Our photometric follow-up campaign of SN 2013gc spans 7 years, starting from March 2010. The first detection of the SN is on 2013 August 26, $\sim$1.5 months before the discovery announcement. With only one observation during the rising phase, it is not possible to precisely determine the explosion epoch. The last non-detection is dated 2013 June 19, hence the SN must be exploded between the above two dates. As we will detail in Sect. \ref{discussion}, we assume 2013 August 16 (MJD 56520) as explosion date.
The object has been followed for about 2 years, up to June 2015. Pre-explosion images and those obtained at very late phases are unfiltered. The multi-band optical magnitudes are reported in Tab. \ref{tab5} and the infrared (IR) in Tab. \ref{tab7}, while unfiltered magnitudes are listed in Tab. \ref{tab6}.

\begin{figure*}
\includegraphics[width=2.1\columnwidth]{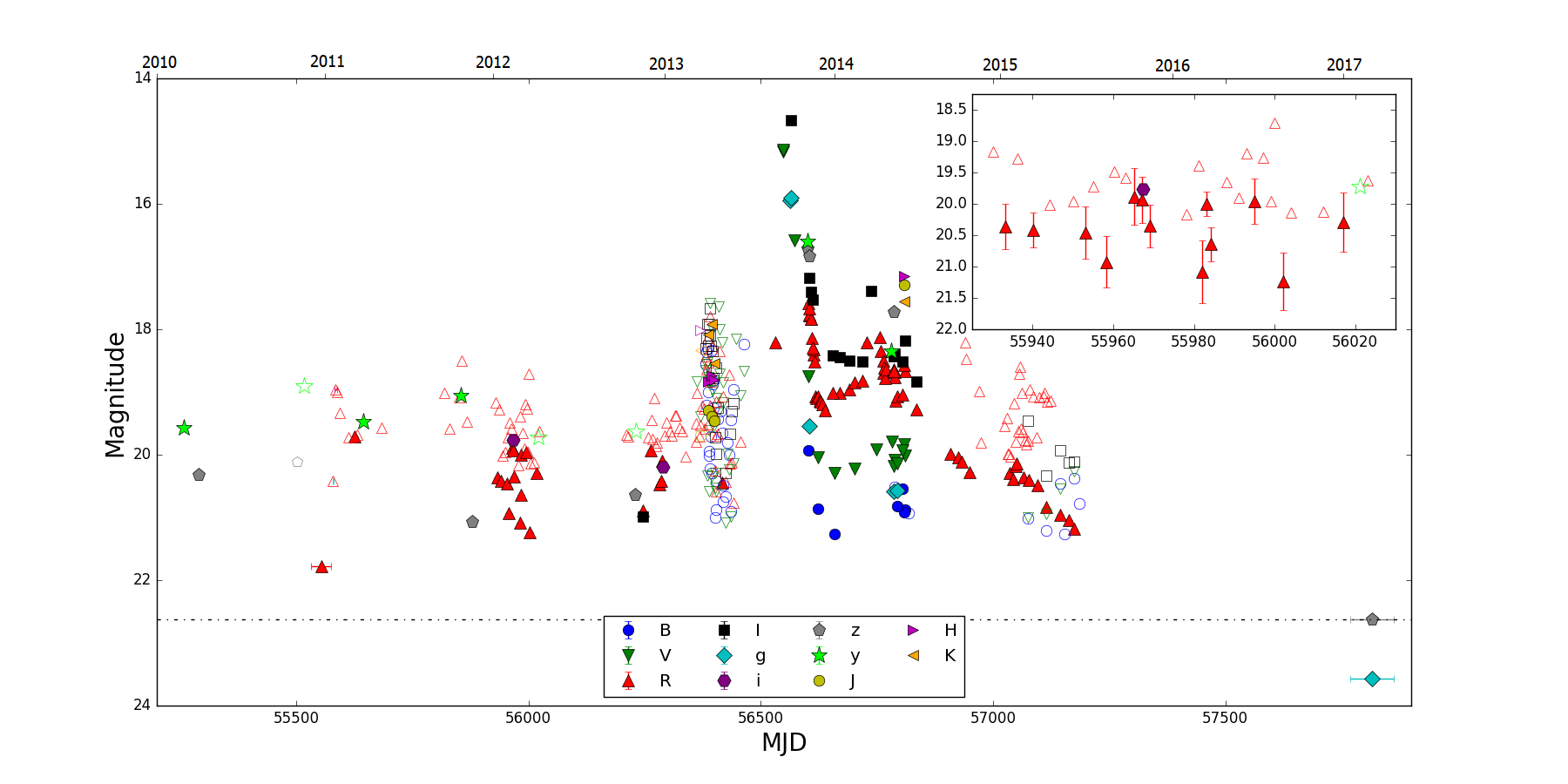} \\
\includegraphics[width=1.3\columnwidth]{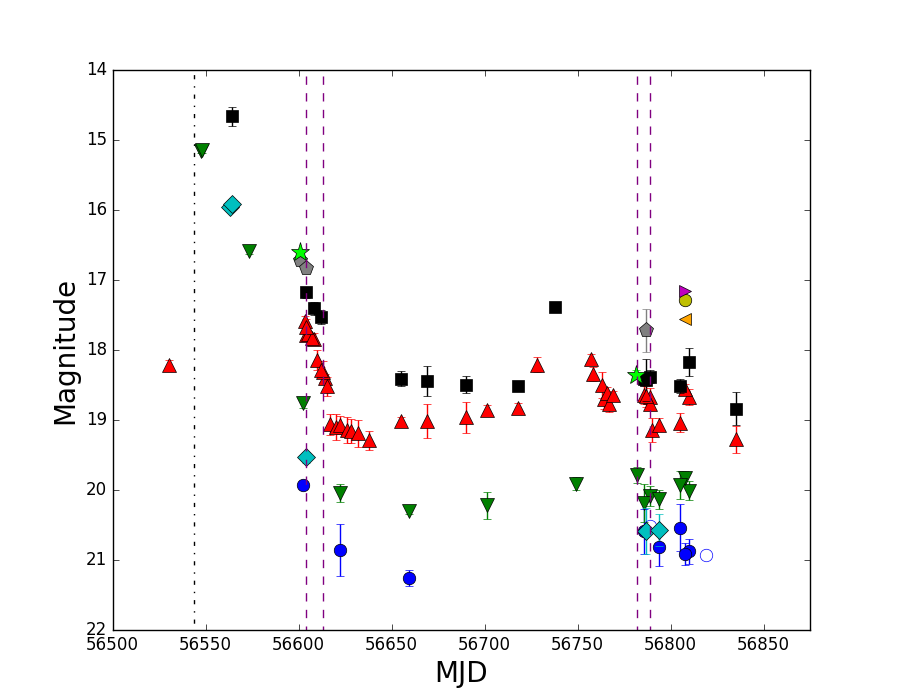}
\caption{Optical and NIR light curves of SN 2013gc. Top: the full 7-year coverage. For sake of clearness, error bars are not reported here. A blow-up on the light curve of the transient during the 2012 variability period (see Sect. \ref{historia}) is shown in the top-right inset. The horizontal line indicates the magnitude of the brightest detection of 2017, showing that the object has faded with respect to the pre-SN eruptive-phase. Bottom: blow-up on the post-explosion evolution. The purple, dashed lines mark the epochs of the available spectra, while the black dot-dashed line marks the date of the assumed maximum. Upper limits are indicated as empty symbols. For each band, a different symbol is used.}
\label{lightcurve}
\end{figure*}

The photometric coverage of the SN evolution in the \textit{BVRIJHK} and \textit{gzy} bands is shown in Fig. \ref{lightcurve}. Magnitudes are not corrected for the line of sight extinction. We assume MJD 56544$\pm7$ as an indicative epoch for the maximum luminosity, that is 4 days before the brightest $V$ band magnitude (MJD 56548) and 20 days before the brightest point in the $I$ band (MJD 56564).
This agrees with the estimate of \citet{2013CBET.3699....1A}.

The light curve shows a fast linear decline after maximum, with a rate of $6.6\pm0.2$, $7.5\pm0.3$ and $6.1\pm0.2$ mag (100 d)$^{-1}$ in the $V$, $R$ and $I$ bands, respectively. Such decline rates are typical of the `Linear' subclass of type II SNe \citep{2016MNRAS.459.3939V}. 
Two EFOSC2 observations have been performed 3 and 4 days after our estimated maximum, when the SN was at $V$ = 15.1 mag (hence, $M_V\sim-16.6$ mag). 

The initial linear decline is followed by a slower decline in all bands. This phase, lasting $\sim$20 days, is well sampled in the $R$ band with a slope of $1.02\pm0.11$ mag (100 d)$^{-1}$. 
This decline slope is consistent with that expected from the decay rate of $^{56}$Co into $^{56}$Fe. The second decline ends at phase $\sim$90 days.
This short fraction of the light curve will be used to guess an upper limit on the $^{56}$Ni mass ejected by SN 2013gc (see Sect. \ref{discussion}). However, we will note in Sect. \ref{spectra} that the CSM/ejecta interaction features are present at all phases of the SN evolution.

Later on, the light curve shows a sort of plateau at around 20.5, 19.0 and 18.5 mag in the $V$, $R$ and $I$ bands, respectively. The plateau lasts about 70 days. We believe that the abrupt stop of the luminosity decline at +120~d marks the onset of a new, stronger ejecta-CSM interaction episode, that hereafter will become the primary source of energy for the SN. Later on, a rebrightening in the $R$ and $I$ bands is observed, with a rise of 1 mag. A secondary peak in $I$-band is reached on MJD 56738 (+194 days). Then, the light curve rapidly fades.

From +300 to +630 days, the light curve presents a well-defined linear decay, with a slope of $0.43\pm0.05$ mag (100 d)$^{-1}$, flatter than the decay rate of $^{56}$Co. However, the SN is detected only in the $R$ band. Around 2 years after the explosion, due to the faintness of the object, the photometric measurements are very close to the detection limits.

The SN field was imaged again by the DECam Plane Survey (DECaPS, PI Rau), with the 4-meter `V. Blanco' telescope at CTIO, between January and April 2017. From this survey, we collect 3 stacked images in the Sloan $grz$-bands. The images are calibrated using the PANSTARRS DR1 catalog. We detect a faint source at the position of SN 2013gc at $g=23.57\pm0.22$ mag, $r=22.63\pm0.18$ mag, $z=22.62\pm0.24$ mag. These detections are over 1 mag fainter than those of the outbursts observed in 2010 to 2013, and this can be used as an argument to support the terminal explosion of the progenitor star. 

\section{Colour and absolute light curves: comparison with similar objects}
\label{colour}
\subsection{Colour curves}
We compare the $(B-V)_0$, $(V-R)_0$ and $(R-I)_0$ colour curves of SN 2013gc with those of type IId SN 1994aj \citep{1998MNRAS.294..448B}, SN 1996L \citep{1999MNRAS.305..811B}, SN 1996al \citep{2016MNRAS.456.3296B} and SN 2000P in Fig. \ref{color}. The photometry of SN 2000P is published in this paper for the first time, and is reported in appendix \ref{2000p}. For SN 1996al, we used the reddening and the distance reported in \citet{2016MNRAS.456.3296B} ($\mu=31.80\pm0.2$ mag, $E(B-V)=0.11\pm0.05$ mag), while for SN 1994aj, SN 1996L and SN 2000P we adopted the reddening and the luminosity distances reported in NED for the respective host galaxies (SN 1994aj: $\mu=35.72$ mag, $A_V=0.115$ mag; SN 1996L: $\mu=35.76$ mag, $A_V=0.255$ mag; SN 2000P: $\mu=32.75$ mag, $A_V=0.209$ mag). All values are obtained with the assumption $H_0=73$ km s$^{-1}$Mpc$^{-1}$.

Soon after maximum, the objects have colour indices around 0 mag. At early phases the $(B-V)_0$ and $(V-R)_0$ colour curves of our SN IId sample shows an evolution to redder colours faster up to about 60 days, then slower until 150 days, from $\sim$0 to $\sim$1 mag. $(R-I)_0$ for SN 1996al, SN 1996L, SN 2000P and SN 2013gc remains almost flat.
At late phases (>240 days), $(B-V)_0$ increases rapidly (30 days) from 0 to 0.7 mag in SN 2013gc, as expected from a cooling photosphere (see Fig. \ref{color}). Meanwhile, $(V-R)_0$ has not a monotonic trend, showing significant oscillations of 0.3 mag at later phases, whilst $(R-I)_0$ has a flatter evolution, stabilizing around 0 mag.
\begin{figure}
\includegraphics[width=1.1\columnwidth]{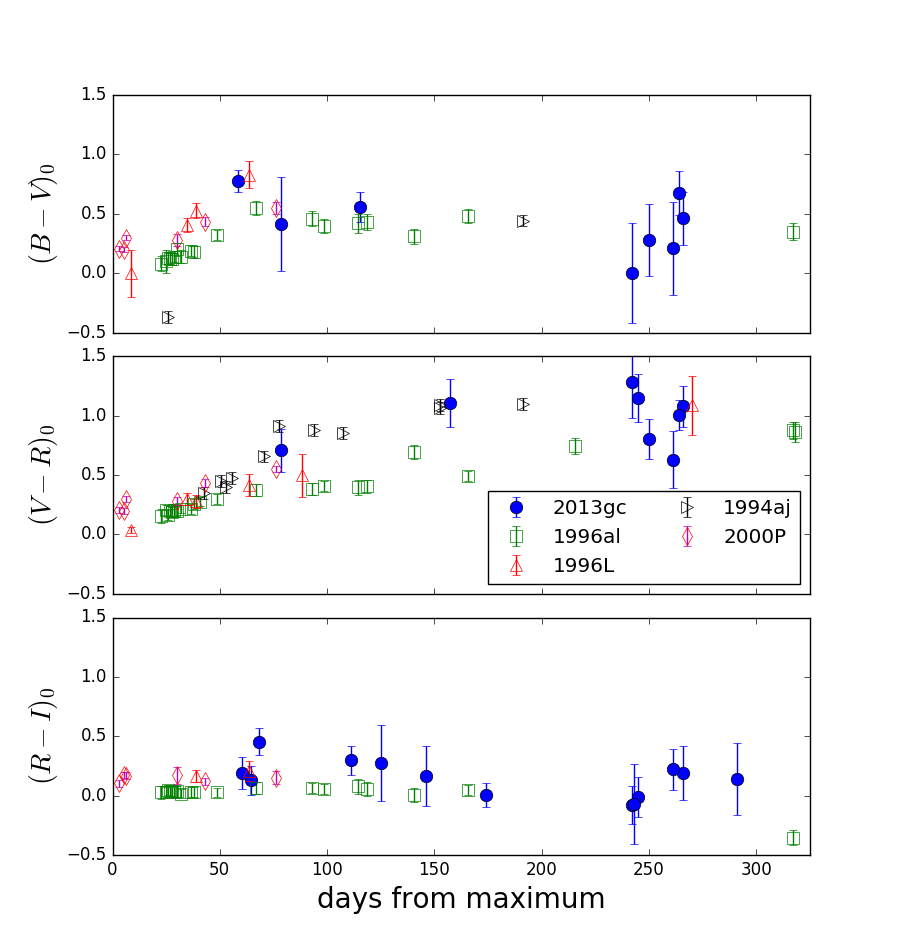}
\caption{Colour curves of SN 2013gc (blue circles), corrected for extincion, shown with those of the comparison SNe 1996al (green squares), 1996L (red triangles), 1994aj (black triangles) and 2000P (purple diamonds). The phase of SN 2000P is from the discovery date, not from the maximum, that probably was not observed (Fig. \ref{2000P} in appendix). $(B-V)_0$ is in the upper panel, $(V-R)_0$ in the middle and $(R-I)_0$ in the bottom panel. Error bars are reported for all objects.}
\label{color}
\end{figure}

\subsection{Absolute light curves}
With the distance adopted in Sect. \ref{discover} and the extinction reported in Tab. \ref{tab1} for SN 2013gc, we obtain $M_I\sim-16.5\pm0.25$ mag for the brightest observation (without accounting for the error on the reddening and on the host galaxy extinction). This value is close to the mean luminosity of `normal' SNe IIL $ \langle M_B \rangle =-16.8\pm0.5$ \citep{2016MNRAS.459.3939V} but, as discussed before, our brightest detection is not coincident with the real maximum. 
\begin{figure}
\begin{center}
\includegraphics[width=1.1\columnwidth]{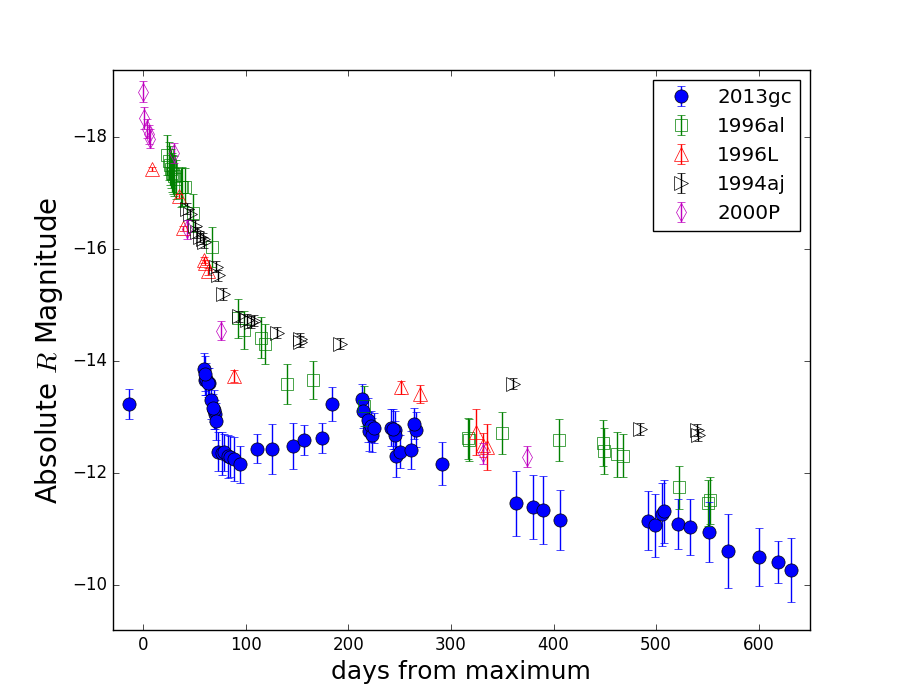}
\caption{Absolute $R$-band light curves of SN 2013gc (blue circles) and comparison objects. Error bars are reported. The different symbols indicate the same SNe as in Fig. \ref{color}.}
\label{absolute}
\end{center}
\end{figure}

The $R$-band absolute light curves of SN 2013gc and the type IId SNe 1996al, 1996L, 1994aj and 2000P are compared in Fig. \ref{absolute}.
From this comparison, we note that the whole absolute light curve of SN 2013gc is fainter than those of comparison SNe by about 2 mag. However, the decline (between 40 and 70 days after maximum) of SN 2013gc is quite similar to the average of the sample.

Because of the low Galactic latitude of the host galaxy, the Milky Way extinction is uncertain.
Shifting upwards SN 2013gc by 2 mag, a fair agreement between SN 2013gc and  other SNe IId would be found, with the exception of the late rebrightening of 1 mag (lasting 40 days) which remains a distinctive feature of SN 2013gc.

The source detected in the DECam images in 2017, with $M_r\approx -8.9$ mag and an intrinsic colour $(g-r)_0 \approx$ 0.5 mag, can be a contaminant background H II region or even a residual signature of the SN. 

\section{Analysis of pre-SN data}
\label{historia}
The SN field was sparsely monitored for more than 3 years before the explosion. We found several archival images taken between 2010 and early 2013 with a detection of a source at the SN position, although in some cases with 
large error bars. This is very likely an indication that the SN 2013gc progenitor experienced a long-lasting luminosity variability, most likely a major eruption started a few years before the SN explosion.

The PS1 survey sparsely monitored the field between 2010 and 2013. The first PS1 images of March 2010, in the $z$ and $y$ bands, show a source with a magnitude of about 20. Then, a stack of the nine best-quality frames taken between 2010 December 2 and 2011 January 14 from the PTF survey (PI Rau) has been produced. In this deep image, we detect a very faint source at $R\sim$21.8 mag, which is close the bona-fide limiting magnitude of the survey. The absolute magnitude of the source is only $M_R\sim-9.65$ mag.
From the same survey, we combine the images obtained from 2011 January 17 to 19, and from 2011 January 21 to 27. The seeing of those nights was quite poor (3 arcsec), and only upper limits are obtained.

We found a few images taken later in 2011, with only one detection on 2011 March 06 (MJD 55626). For this event we estimate $M_R=-11.74\pm$0.35 mag, although we cannot constrain the duration of the outburst. In fact, in frames obtained 12 days before and 6 days after the burst, nothing is visible at the same limiting magnitude.

On early 2012, the field was well monitored, and a few additional sparse detections are found (see the top-right box in Fig. \ref{lightcurve}). This provides additional evidence of the long-lasting photometric instability of the SN progenitor. The faintest detections are at 21 mag, corresponding to $M_R\sim-10.5$ mag, but in other cases the source has an apparent mag between 20 and 21. The source has also been detected by PS1 on 2011 October 20 in the $y$ band, at 19 mag, which is almost 3 mag brighter than the faintest PTF detection measured 9-10 months before.
We now describe a sequence of events: on 2012 February 25 (MJD 55982), the source is hardly visible. The day after, the object is in flare, showing a brightening of 1 mag. Then, on 2012 February 27, the transient has faded again by 0.5 mag (inset of Fig. \ref{lightcurve}). This sequence highlights a scenario with a short-duration flares occurring during a long eruptive phase. The absolute magnitude estimated for this flare is $R=-11.41\pm$0.40 mag, similar to that observed in sole LBV-like outburst, such as SN 2000ch \citep{2004PASP..116..326W}. The upper limits measured in a number of lower-quality images are not very deep (<20 mag), hence they are not very constraining.

After some months without images due to the heliac conjunction, the monitoring campaign restarted in late 2012, and the object was detected in several frames. In one of them, on 2012 December 1, the object reached $M_R=-11.51\pm0.50$ mag.
In March 2013, the explosion of SN 2013ak in the same galaxy triggered a vast observational campaign, and tens of multi-band images of the field are hence available. In one of those images, taken with the Gemini South telescope, a brightening is clearly revealed on 2013 May 4 at $R=20.45$ mag ($M_R\sim-11.00\pm0.23$ mag). Also this event is probably a short-duration flare. In fact, the source is not detected in images taken one day before and one day after. However, the quality of these images, taken with the PROMPT telescopes, is lower than the image of the Gemini South telescope.
We remark that the peak luminosity of the four impostor events described above is very similar.
After June 2013, the region was again in heliac conjunction. Then, when the field became again visible at around mid-August 2013, the SN was already visible.\\

\begin{table*}
\caption{Basic information of the 5 spectra of SN 2013gc. The phase is with respect to the adopted maximum (MJD 56544). To evaluate the instrumental resolution, we measured the FWHM of the [\ion{O}{I}] night sky lines.}
\label{tab2}
\begin{tabular}{lcllccl}
\hline
Date & MJD & Phase & Instrument & sky lines & spectral range & exp. times \\
& & (d) & & FWHM (\AA) & (\AA) & (s) \\
\hline
2013 November 8  & 56604.20 & +60  & WFCCD+blue grism 	  & 8.2 & 3620-9180 & 2$\times$900 \\
2013 November 17 & 56613.17 & +69  & Goodman Spectrograph & 5.0 & 3765-8830 & 2$\times$3600\\
2014 May 5 		 & 56782.09 & +238 & WFCCD+blue grism 	  & 7.7 & 3630-9200 & 2$\times$1000\\
2014 May 11		 & 56788.98 & +245 & Goodman Spectrograph & 7.6 & 4050-8940 & 2$\times$2700\\
2014 May 12		 & 56789.05 & +245 & Goodman Spectrograph & 1.7 & 6250-7500 & 2700 \\
\hline
\end{tabular}
\end{table*}

\begin{figure*}
\includegraphics[width=2.25\columnwidth]{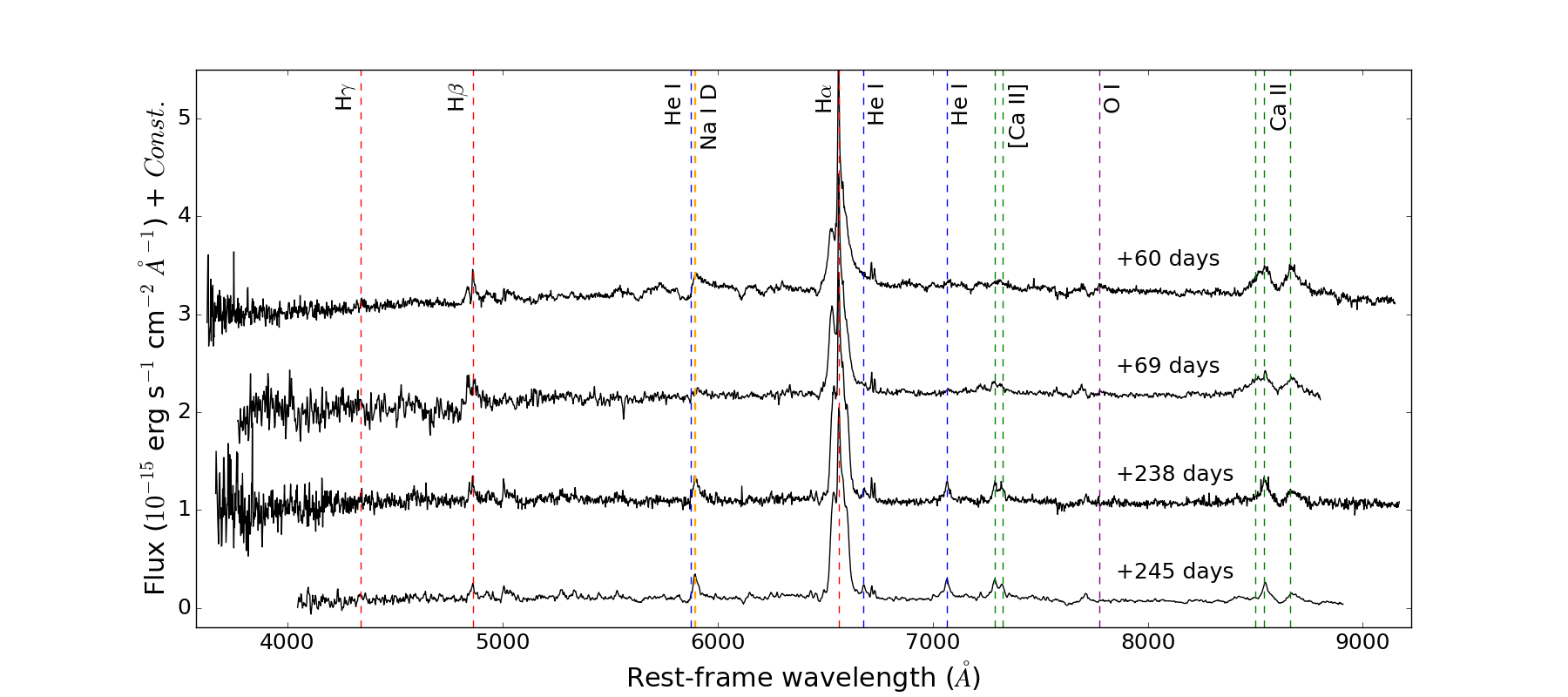}
\includegraphics[width=1.8\columnwidth]{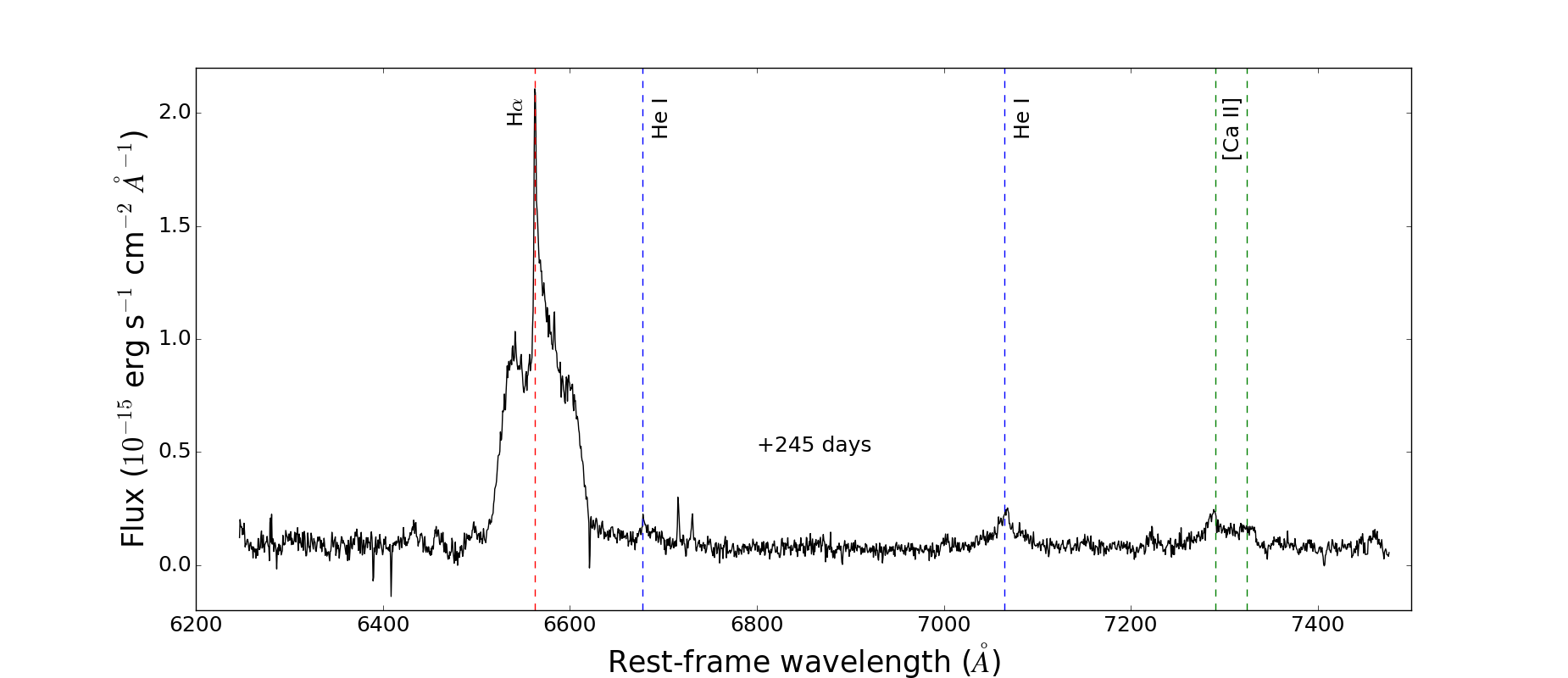}
\caption{Top: the 4 low-resolution spectra of SN 2013gc, corrected for redshift and reddening. Bottom: The mid-resolution spectrum of SN 2013gc, dated 2014 May 12. The spectrum covers 1200 \AA\quad around the H$\alpha$ line. The most prominent lines are identified in figure.} 
\label{spettri}
\end{figure*}

\section{Spectroscopy}
\label{spectra}
We obtained 5 optical spectra of SN 2013gc with the following two instrumental configurations:
\begin{itemize}
\item The 4.1-meter ``SOAR'' telescope at CTIO, with the Goodman Spectrograph\footnote{It has a 7.2 arcmin diameter FOV, with a 0.15 arcsec/pixel scale. We used the Red Camera. The Spectrograph can also do imaging. The Red Camera is equipped with a 4096$\times$4112 pixel, back-illuminated, E2V CCD.
(http://www.ctio.noao.edu/soar/content/goodman-red-camera)} (PI F. Bufano).

\item The Du Pont telescope at Las Campanas Observatory with the WFCCD/WF4K\footnote{The camera covers a 25' field, with a scale of 0.484 "/pixel. The detector has 4064$\times$4064 pixels. (http://www.lco.cl/telescopes-information/irenee-du-pont/instruments/)} (PI P. Lira).
\end{itemize}
The pointing images of these instrument were also used for photometry.
The first spectrum, taken one day after the discovery, was used for the spectral classification. The second spectrum is dated 2013 November 17, during the first dimming phase. The last two spectra were obtained after the second luminosity peak, on 2014 May 5 and May 12. On 2014 May 12 a medium resolution spectrum around the H$\alpha$ region was also taken. Technical details of the five spectra are reported in Tab. \ref{tab2}.

\subsection{Spectroscopic reduction and line identification}
All spectra were pre-reduced and calibrated using routine IRAF packages. The 2-dimensional frames were corrected for bias and flat-field, then the 1-dimensional spectra were extracted, and sky lines and cosmic rays were removed. On that spectrum, we performed wavelength and flux calibrations, using arc lamps and spectrophotometric standard stars. The spectral fluxes were scaled according to the $R$-band photometry of the nearest night. The spectra were also corrected for the strongest telluric absorption bands.
The five spectra were corrected for redshift and reddening, adopting the values of $z$ and $A_V$ given in Tab. \ref{tab1}.
The calibrated low-resolution spectra, with line identification, are shown in Fig. \ref{spettri}, while the medium-resolution spectrum is plotted in the bottom panel.

A weak, red continuum ($T_{bb}\sim$4000 K) is present in the early spectra. Absorption features of metals, and Balmer lines in emission are also observed. 
H$\alpha$ is the most prominent feature in all spectra, and its profile is described in detail in Sect. \ref{profile}. H$\beta$ has a similar profile as H$\alpha$. H$\gamma$ is not clearly detected, although the spectra have a low SNR at the blue wavelengths.
He lines are weak in early spectra, but become more intense later. \ion{He}{I} $\lambda$7065 is the stongest line, followed by $\lambda$6678 and $\lambda$4922.
\ion{He}{I} 5876 is blended with the \ion{Na}{I}D doublet in emission.

To evaluate the evolution of the He lines, we measure the flux ratio H$\alpha$/\ion{He}{I} $\lambda$7065 in all spectra. We choose the \ion{He}{I} $\lambda$7065 line because it is isolated and not significantly blended with other lines.
In the first two spectra, the ratio is around $90\pm10$, but the flux measurement of the He line is difficult due to its faintness. In the +238 d spectrum, the ratio is lowered to $\sim$30, and in the +245 d spectra is only $18\pm2$.

We identify \ion{Fe}{II} features (multiplets 40, 42, 46, 48, 49, 199).
The \ion{Fe}{II} (46) $\lambda$6113 line is observed in absorption at early phases, while it is in emission at late epochs.

At late phases, the flux contribution of the spectral continuum is negligible, and the line profiles have changed. We identify the \ion{Ca}{II} IR triplet lines, detected as broad emission. 
The FWHM of the \ion{Ca}{II} $\lambda$ 8662 line is 3400$\pm$100 km s$^{-1}$. There is a possible detection of \ion{O}{I} $\lambda$7774 in the first spectrum.
More in general, we note an increasing number of emission lines, especially from forbidden transitions.
In particular, we identify: [\ion{N}{II}] $\lambda\lambda$ 6548,6584 lines (although blended with the H$\alpha$ line), [\ion{S}{II}] $\lambda\lambda$ 6716,6731 lines, [\ion{Ca}{II}] $\lambda\lambda$ 7291,7324 lines. Narrow [\ion{O}{III}] $\lambda\lambda$ 4959,5007 lines are superimposed on broader \ion{Fe}{II} feature emission.
Narrow [\ion{S}{II}], [\ion{N}{II}] and [\ion{O}{III}] lines likely arise from unresolved background contamination.
In fact, in a 2012 image taken by the VST telescope equipped with OMEGACAM and a narrow H$\alpha$ filter during the VPHAS+ survey (PI Drew), a diffuse, elongated source is visible. This object can be a foreground H II region. In contrast with other forbidden lines, [\ion{Ca}{II}] lines are attibuted to the SN environment, and seem to become stronger with time.

\subsection{Comparison with similar objects}
\label{ha}
Our spectra are compared with those of SN 1996al, SN 1996L, SN 1994aj and SN 2000P obtained at around the same phases. All spectra of SN 2000P are presented in this paper for the first time (see Appendix \ref{2000p}).
\begin{figure*}
\includegraphics[width=2.25\columnwidth]{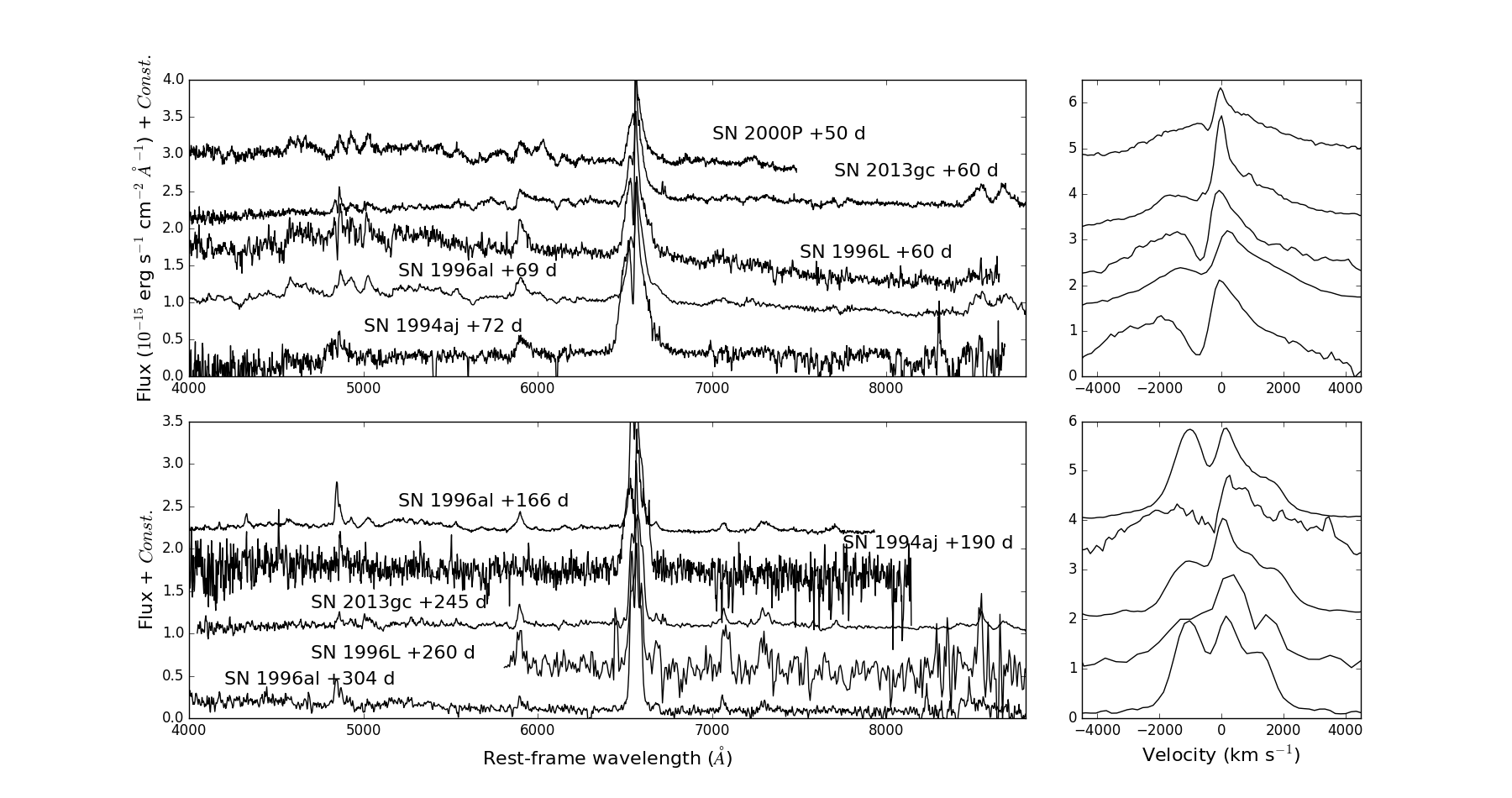}
\label{8}
\caption{Comparison of the +60 (top) and +245 (bottom) days spectra of SN 2013gc with those of reference SNe IId at approximately the same phase. On the right panel, the same comparison is done for H$\alpha$ profiles at the same epochs.}
\end{figure*}

The early spectra of SN 2013gc are similar to those of SN 1994aj, SN 1996al and SN 1996L taken at 60-70 days after maximum, in particular with respect to the H$\alpha$ profile and the \ion{Na}{I} D P-Cygni line. In all these SNe, H$\alpha$ and H$\beta$ show a narrow P-Cygni absorption over a broader P-Cygni component. The main difference at early times is that SN 1996L have a hotter continuum, peaking around 5000 \AA, giving a black-body temperature $T_{bb}$ of $\sim$6000 K.

One of the most interesting comparison is with SN 2000P, which has an excellent spectroscopic dataset. As for SN 2013gc, H$\alpha$ is characterized by a very narrow P-Cygni absorption and an extended Lorentzian red wing. H$\beta$ has a similar narrow P-Cygni feature. Many bumps, likely due to \ion{Fe}{II}, are observed in the spectra of both objects. 
It is worth comparing the spectra and the H$\alpha$ profile evolution of SN 2000P (Fig. \ref{seq}) with those of SN 1996al \citep[Fig. 5 and 7 of][]{2016MNRAS.456.3296B}. In the spectra of SN 2000P obtained a very few days after the maximum light, the blue continuum indicates a high gas temperature. H$\alpha$ and H$\beta$ show a double P-Cygni feature, along with a broad emission from \ion{He}{I} $\lambda$5876 and \ion{Na}{I} D lines. At phase +30 days, many bumps from metals arise and the \ion{Ca}{II} triplet becomes well visible, similar features are visible in the first spectrum of SN 2013gc. From this epoch, a blue-shifted H$\alpha$ bump starts to develop at an intermediate-width, and grows in strength with time. One year after the maximum, H$\alpha$ is nearly the only observable feature, and the blue-shifted component is more prominent than that at the rest frame. 

The late-time spectra of SN 2013gc and SN 2000P are not sufficiently close in phase to make a reasonable comparison. 
A comparison of the spectra of SN 2013gc with other SNe IId at similar phases is provided in Fig. 6.
In the comparison of early-time spectra, we note that H$\beta$ is quite prominent, with the Balmer decrement H$\alpha$/H$\beta$ being higher in SN 2013gc than in other SNe IId.
In particular, for SN 1996al at $\sim$ +60 days, \citet{2016MNRAS.456.3296B} found an H$\alpha$/H$\beta$ ratio of 5, while in SN 2013gc we infer a ratio of around 10. 
In the late spectra the ratio increases to $\sim$20. This comparison with SN~1996al supports the possibility that the reddening in the direction of ESO 430-20 may have been underestimated.

\begin{figure}
\includegraphics[width=0.95\columnwidth]{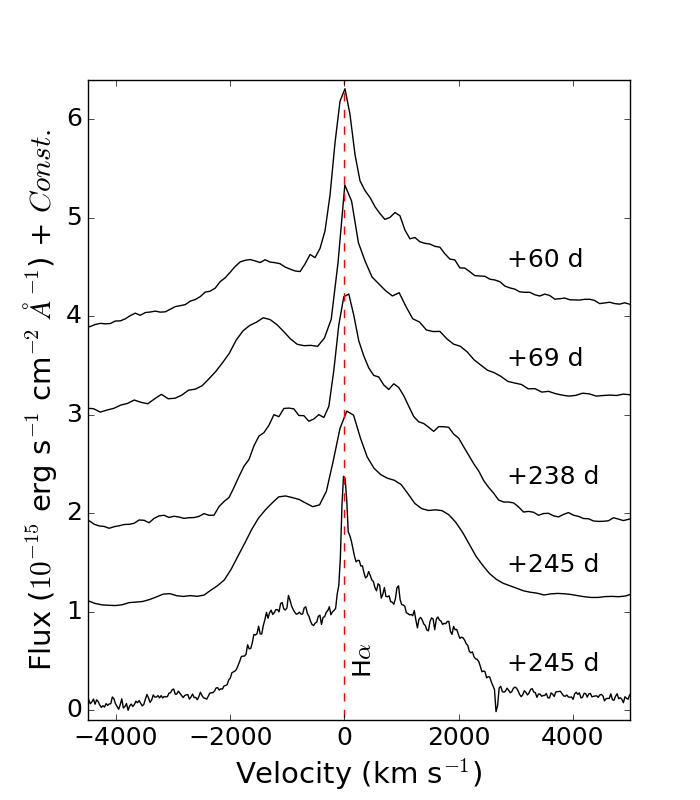}
\caption{Blow up of H$\alpha$ line in the 5 available spectra. Velocities with respect to the rest frame are reported in abscissa. The zero velocity is marked with a dotted line.}
\label{halfa}
\end{figure}

\subsection{The H$\alpha$ profile}
\label{profile}
\begin{figure*}
\includegraphics[width=\columnwidth]{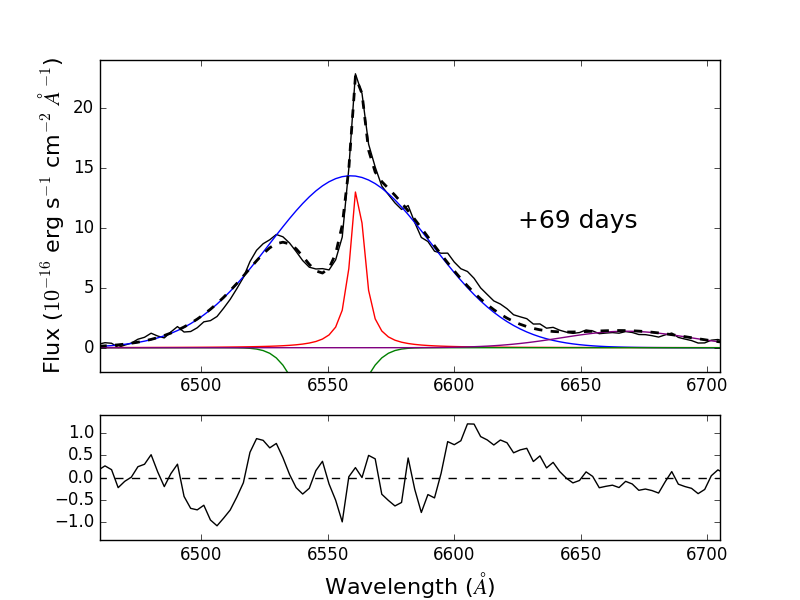}
\includegraphics[width=\columnwidth]{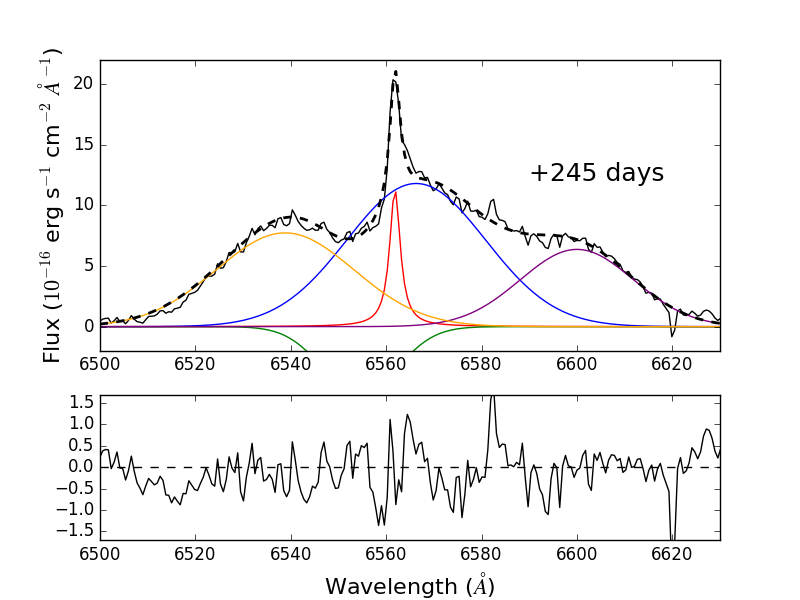}
\caption{Decomposition of the H$\alpha$ profile through multiple components. Left: In the +69 days spectrum the profile is well fitted using a broad Gaussian emission (blue), on top of which a narrow Lorentzian emission (red) stands, and a Gaussian absorption (green) for the P-Cygni. The bump due to the \ion{He}{I} $\lambda$6678 is also fitted with a Gaussian emission (purple).
Right: In the mid-resolution (FWHM=1.7 \AA) +245 days spectrum, the H$\alpha$ profile is decomposed with an intermediate-width Gaussian emission (blue), a narrow Lorentzian emission (red) and a Gaussian absorption for the P-Cygni (green). Two additional gaussian components are required, red-shifted (purple) and blue-shifted (orange).
In both graphs, the sum of the 4 components is shown with a black dashed line. Residuals between the data and the fit are reported on the bottom panels.}
\label{deconvol}
\end{figure*}

\begin{table*}
\caption{FWHM and central wavelengths of the various components of the H$\alpha$ profile in all spectra. For the P-Cygni component, the velocity of the minimum with respect to the rest-frame is reported.}
\label{tab3}
\begin{tabular}{lccccc}
\hline
Phase & (Em)$_{broad}$ & (Em)$_{narrow}$ & (Ab)$_{P-Cygni}$ & Red Wing & Blue Wing \\
\hline
& v$_{FWHM}$; $\lambda_c$ & v$_{FWHM}$; $\lambda_c$ & Min. vel. & v$_{FWHM}$; $\lambda_c$ & v$_{FWHM}$; $\lambda_c$\\
(d) & (km s$^{-1}$);(\AA) & (km s$^{-1}$);(\AA) & (km s$^{-1}$) & (km s$^{-1}$);(\AA) & (km s$^{-1}$);(\AA) \\
\hline
+60  & 3400$\pm$100; 6559 & 300$\pm$15; 6561.7 & 560 & & \\
+69  & 3450$\pm$100; 6557 & 340$\pm$20; 6561.7 & 470 & & \\
+238 & 1600$\pm$50; 6565  & 420$\pm$20; 6561.2 & 420 & 1800$\pm$50; 6598  & 1600$\pm$50; 6540 \\
+245 & 1600$\pm$50; 6566  & 350$\pm$20; 6563.5 & 460 & 1650$\pm$50; 6598  & 1600$\pm$50; 6542 \\
+245 & 1550$\pm$50; 6566  & 120$\pm$5;  6561.8 & 380 & 1300$\pm$30; 6600  & 1550$\pm$50; 6539 \\
\hline
\end{tabular}
\end{table*}

The analysis of the H$\alpha$ profile can provide information on the CSM around the SN. The H$\alpha$ profile is complex, and consists of multiple components (see Fig. \ref{halfa}). In the early spectra, a narrow P-Cygni absorption is superposed on a broad component. The line is asymmetric, with an extended red wing.
The simultaneous presence of a broad component from the fast ejecta and a narrow P-Cygni profile from slow circumstellar wind is a characterizing feature of type IId SN spectra.
The H$\alpha$  profile shows an evolution from the early  (+60 d) to the late (+245 d) epochs, with the velocities of the broader components progressively decreasing with time.
In order to identify the different line components, we deblended the H$\alpha$ profile, following \citet{2016MNRAS.456.3296B}. We considered the spectra obtained +69 and +245 days. 
The results are illustrated in Fig. \ref{deconvol}, and the velocities of the different line components are reported in Tab. \ref{tab3}.

At the early epoch, the relatively broad H$\alpha$ component has a velocity of 3400 km s$^{-1}$, which is very similar to the FWHM velocity ($v_{FWHM}$) of the \ion{Ca}{II} $\lambda$8662 line. The FWHM of this component can be considered representative of the SN ejecta velocity.
In the late spectrum, H$\alpha$ has been deblended using 3 distinct intermediate-width emission components, with comparable $v_{FWHM}$ ranging between 1300 and 1800 km s$^{-1}$. These components correspond to 3 distinct emitting regions: a first one (blue shifted from the rest-wavelength) moving towards the observer, a second (red shifted) one produced by receding gas, and a third one, centered at the rest-frame.

In principle, from the velocities of the ejecta and the CSM, one can infer the time of ejection of the CSM which later interacts with the SN ejecta. We adopt the velocities derived from the first spectrum, because it is the closest available to the explosion. We adopt 3400 km s$^{-1}$ for the ejecta (from the FWHM of the broad component), and 560 km s$^{-1}$ for the wind (from the minimum of the narrow P-Cygni component). Although the velocity at the explosion should be used for freely expanding ejecta, those of SN 2013gc are likely shocked already soon after the explosion, hence the value inferred from the first spectrum is a fair approximation of the ejecta velocity. The explosion epoch is unknown, but it is constrained between MJD 56463 (the last non-detection) and MJD 56530 (the first SN detection). 
Assuming that the collision of the SN ejecta with a dense circumstellar shell marks the onset of the light curve plateau (MJD 56640$\pm10$), the material would have been ejected between MJD $\sim$55560 (December 2010-January 2011) and MJD $\sim$55970 (February 2012) for the above two constraints on the explosion epoch, respectively. On the other hand, at about MJD 56720$\pm10$, we note a major brightening of the light curve, which is an evidence of enhanced CSM-ejecta interaction. With the same velocities and the explosion time interval, the second shell would have been ejected between MJD $\sim$55150 (November 2009) and MJD $\sim$55560 (December 2010-January 2011).
The above constraints on the mass loss epochs are consistent with the timing of the pre-SN detections, and favour an explosion occurring soon after the last non-detection (Sect. \ref{historia} and Fig. \ref{lightcurve}, top panel). From March 2010, we directly witnessed the outbursts that produced the SN CSM and determined the type IIn/IId observables.

\section{Discussion}
\label{discussion}
The composite Balmer line profiles, with the simultaneous presence of broad and narrow components with P Cygni profiles, makes SN 2013gc a member of the IId sub-class of type IIn SNe \citep{2000MmSAI..71..323B}. This fact is supported by the comparison of the colour and absolute light curves of SN 2013gc with those of known SNe IId which show a similar evolution. In particular, the onset of strong ejecta-CSM interaction in SN 2013gc (between 120 and 140 days after the explosion), is compatible with that observed in similar SNe \citep[typically between 100 and 150 days,][]{2000MmSAI..71..323B}.

\begin{figure}
\includegraphics[width=1.1\columnwidth]{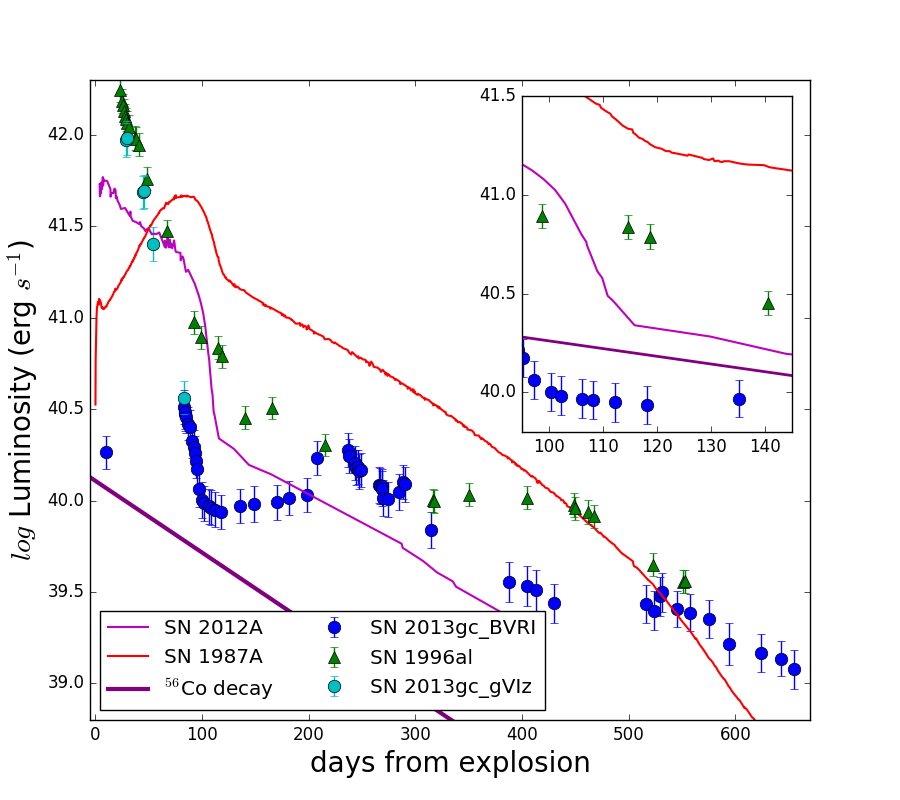}
\caption{Quasi-bolometric $BVRI$ light curves of SNe 1987A, 1996al, 2012A and 2013gc. The inset shows a blow-up of the region between 95 and 145 days after the explosion, in which there is evidence of $^{56}$Co radioactive decay. The decay slope of $^{56}$Co (0.98 mag (100 d)$^{-1}$) is reported with a purple line for comparison. The early phase (\textit{gVIz}) bolometric light curve is plotted with cyan dots.}
\label{bolom}
\end{figure}

\subsection{Bolometric luminosity and the $^{56}$Ni mass}
We calculated the pseudo-bolometric light curve of SN 2013gc accounting for the contribution in the \textit{BVRI} bands only. For epochs without observations in some bands, we made an interpolation to the available data using the $R$-band light curve as reference, and assuming a constant colour index.
The maximum is not covered by $R$-band images hence, to constrain the peak luminosity, we calculated the pseudo-bolometric light curve accounting only for the \textit{gVIz} band contribution. We assumed the $V-I$ and $V-z$ colours to be constant between 30 and 80 days after the explosion, at 1.5 and 2.0 mag, respectively, as derived from the closest photometry of 2013 November 7.
We fixed the explosion date on MJD 56520$\pm$7, about 10 days before the first SN detection. This implies a rise-time of 23$\pm$7 days from the explosion to the maximum. Then a first steep decline is observed. In order to estimate an upper limit to the ejected $^{56}$Ni mass, we focus on the light curve portion having a decline slope similar to that of $^{56}$Co. We identify a very short time interval during which the decline slope is compatible with the $^{56}$Co decay rate, i.e. between 97 and 118 days after explosion.

As a comparison, we selected the classical core-collapse SN 1987A, and a type II SN with a shorter plateau, SN 2012A \citep{2013MNRAS.434.1636T}. The choice of SN 2012A was motivated by the fact that the putative $^{56}$Co decay tail started early in SN 2013gc, at $\sim$100 d, when typical type IIp SNe are still in the plateau phase, or during the steep post-plateau luminosity decline. In contrast, SN 2012A was a short-duration IIp event, whose light curve was already following the $^{56}$Co decay at 100~d. Hence, this is a more reliable comparison object to estimate the $^{56}$Ni mass of SN 2013gc. SN 1996al is also considered as a comparison object. The bolometric light curves of the four objects are compared in Fig. \ref{bolom}, with a blow-up on the portion of the SN 2013gc light curve with a decline rate consistent with the $^{56}$Co decay.

The estimated $^{56}$Ni mass of SN 1987A is 0.085 $M_{\odot}$ \citep{2011A&A...532A.100U}, while for SN 2012A the estimate is 0.011$\pm$0.004 $M_{\odot}$ \citep{2013MNRAS.434.1636T}.
We calculated and averaged the bolometric luminosity ratio of SN 2012A and SN 2013gc at three epochs after the explosion, at +112, +118 and +135 days. These three epochs were selected because at the same phases SN 2012A was declining with a rate consistent with the $^{56}$Co decay. The M$^{56}_{Ni}$ is obtained from the following relation:
\begin{equation}
\frac{L_{bol}(2012A)}{L_{bol}(2013gc)}=\dfrac{M^{56}_{Ni}(2012A)}{M^{56}_{Ni}(2013gc)}=2.6^{+3.8}_{-1.1}
\end{equation}

The error bars account also for the uncertainty in the explosion epoch.
From this value, and propagating the errors, the $^{56}$Ni mass estimated for SN 2013gc is $4.2(^{+5.8}_{-3.2})\times10^{-3}M_{\odot}$.
We remark that this value has large uncertainties, because at 135~d the ejecta-CSM interaction becomes very strong, and the light curve becomes flatter than the $^{56}$Co decay. For this reason, the above value of the $^{56}$Ni mass should be regarded as an upper limit. On the other hand, we cannot rule out that an optically thick shell is hiding interior emission, leading us to underestimate the $^{56}$Ni mass.
It is worth noting that for SN 1996al \citet{2016MNRAS.456.3296B} constrained an upper $^{56}$Ni mass limit of 0.018 $M_{\odot}$, a factor of about 4 higher than that inferred for SN 2013gc.

This $^{56}$Ni mass constrained for SN 2013gc is definitely modest, also accounting that other photometric indicators (in particular, the plateau and the second brightening) suggest that the main luminosity powering mechanism is the interaction between the SN ejecta and surrounding material.
For this reason, the $^{56}$Ni mass ejected by SN 2013gc is most likely very low, comparable with that inferred for some under-luminous type II SNe \citep{2004MNRAS.347...74P,2014MNRAS.439.2873S}.

\subsection{The progenitor star} \label{prog}
The interest for SN 2013gc lies in the fact that a prolonged variability phase was observed in the progenitor site before the SN explosion.
So far, only one possible eruptive event was claimed for the progenitor of a SN IId. It was a single detection of a luminous source at the position of SN 1996al in archival H$\alpha$ images obtained 8 years before the SN explosion \citep{2016MNRAS.456.3296B}. For SN 2013gc the indications of pre-SN stellar activity are much more robust, as we witnessed a complex long-lasting eruptive phase prior to the explosion.
As seen in Sect. \ref{historia}, the variability of the progenitor in 2010-2012 resembles those observed in other SN impostors. In particular, the pre-SN stages of SN 2009ip \citep{2013ApJ...767....1P,2013MNRAS.430.1801M} and SNhunt 151 \citep{2017A&A...599A.129T,2018MNRAS.475.2614E}, and the current evolution of SN 2000ch \citep{2010MNRAS.408..181P}.
The bright absolute magnitude of the source in pre-SN images ($M\sim-11$ mag) rules out the possibility that the progenitor was in a quiescent phase.

The deconvolution of the H$\alpha$ profile, in particular from the last spectrum, allows us to reconstruct the structure of the CSM and to constrain the physics of the explosion, and the nature of progenitor star.

The fast initial (50-80 d) drop of the light curve implies small ejected H mass. A modest ejected mass is consistent with two SN scenarios: the low-energy explosion of a relatively low mass progenitor (7-8 M$_\odot$) or, alternatively, a very massive progenitor ($M_{ZAMS}>25$ $M_{\odot}$) ending its existence as a fall-back SN.
Assuming a low-mass progenitor scenario, the pre-SN eruptive phase can be attributed to binary interaction or, alternatively, to a super-AGB phase during which a star loses mass generating a circumstellar cocoon, before exploding as an electron-capture SN \citep[e.g.,][]{1984ApJ...277..791N,2006A&A...450..345K,2009ApJ...705L.138P}.
In the massive progenitor scenario, the star core collapses into a black hole, as most of the stellar mantle falls back onto it. In this way, only the most external layers are ejected. A fall-back SN is also consistent with the low ejected $^{56}$Ni mass \citep{1998ApJ...505..876Z,2010ApJ...719.1445M}, because the inner regions of the star rich in heavy elements remain bound to the collapsed core. A fall-back SN produces a faint explosion. The absence of [\ion{O}{I}] $\lambda\lambda$ 6300,6364 lines in the late spectra is also in agreement with the expectations of the fall-back SN scenario \citep{2009Natur.459..674V}, but can alternatively be explained as due to an optically thick region of interaction, that hides the O-rich ejecta.

High mass stars, such as LBVs and Wolf-Rayer stars, are known to experience severe mass loss via super-winds during the final stages of their life. When they finally produce a core-collapse SN, the material ejected by the SN will interact with the composite circumstellar environment. LBVs are suitable progenitor candidates for types IIn/IId SNe. Such LBV candidates would pass through an eruptive phase with many alleged outbursts, that create a complex and structured CSM. The interaction of the SN ejecta with this CSM would produce unusual features in the light curve, like the plateau and the observed rebrightening.

A third, somewhat different scenario, is a non-terminal explosion event. The precursor activity exhibited by SN 2013gc shares some similarities with the pre-explosion variability of SN 2009ip. The nature of this object and its fate after the two 2012 events, are still somewhat disputed, and a non-terminal mechanism has also been proposed \citep{2013ApJ...767....1P,2014ApJ...780...21M}. This scenario would be consistent with many characteristics of SN 2013gc, including no $^{56}$Ni production, the lack of [\ion{O}{I}] lines and the faint absolute magnitude. In this context, the source detected in the 2017 DECAPS images might be the survived progenitor.As a consequence, the brightest event would have been a SN impostor, that generated a fast moving shell. The following photometric evolution would result from the shell-shell collision. We note, however, that the putative progenitor recovered in the DECAPS images was much fainter than the hystorical detections, still favouring the SN explosion for the 2013 event.

We propose two physical mechanism that possibly driven the precursor variability. Pulsational mass loss in very massive stars, possibly triggered by a Pulsational-Pair Instability, can be accompanied by large luminosity variations \citep{2007Natur.450..390W}. As an alternative, mass loss can be triggered by stellar interaction in an LBV binary system, with eccentric orbits. When the secondary approaches the periastron, the tidal forces can strip away part of the material of the primary, enhancing the mass loss and the luminosity \citep{2010MNRAS.408..181P}. If this is the case, a modulation can be observed in the light curve. The observations do not firmly support one of the two scenarios, because of the modest signal-to-noise of the available photometry, and the lack of adequate cadence.

\subsection{The explosion scenario}
From the decomposition of the H$\alpha$ profile, we can constrain the geometry of the CSM. The presence of multi-component H$\alpha$ line profiles both at early and late-epoch spectra reveals that the CSM-ejecta interaction is likely active from the early phases of the SN evolution, suggesting that progenitor's mass loss has continued until a very short time before the SN explosion. In this context, the above claim is supported by the evidence that the progenitor in outburst was observed until May 2013, only $\sim$100 days before the SN explosion. The light curve plateau and the rebrightening at about 200 days have to be considered as enhanced interaction with a denser CSM regions. 

The H$\alpha$ profile in the late spectra has a composite profile, with three intermediate-width components (with $v_{FWHM}$ exceeding $10^3$ km s$^{-1}$): a blue-shifted component peaking at $-1000$ km s$^{-1}$, a redshifted one centered at +1600 km s$^{-1}$, and a third one in the middle, centered at the rest wavelength of the transition, atop of which a narrow emission is observed with $v_{FWHM}$ of a few $10^2$ km s$^{-1}$. This suggests a complex and structured CSM geometry.

The two components shifted from the rest wavelength reveal a bipolar emitting structure of the CSM, expanding with a core velocity in the range 1300-1800 km s$^{-1}$. The difference of the measured expansion velocities can be explained with an intrinsic difference in the ejection velocity of the two lobes, or possibly by a mismatch between the line of sight and the polar axis of the bipolar nebula. Similar spectral features were observed in the SN IIn 2010jp, although in that case a jet-like explosion in a nearly spherically symmetric CSM was proposed \citep{2012MNRAS.420.1135S}, and the velocities involved were in fact one order of magnitude larger. The kinetic energy of SN~2013gc is much smaller, and is consistent with a fall-back SN scenario (see Sect. \ref{prog}).
A spherically symmetric CSM expanding at much lower velocity is revealed through the presence of a narrow H$\alpha$ component, while a nearly spherical shell, shocked by the interaction, produces the central, intermediate-width component. 
The material shocked by the ejecta is optically thick, and hides (at all epochs) the underlying SN features. 

This CSM configuration shares some similarity with that proposed by \citet{2016MNRAS.456.3296B} for SN 1996al, consisting of an equatorial circumstellar disc producing the H$\alpha$ profile with two emission bumps shifted from the rest wavelength. That material was surrounded by a spherical, clumpy component producing the H$\alpha$ emission at zero velocity. The clumpiness of the SN 1996al CSM was deduced from the evolution of the He lines, that became more prominent with time. 
In analogy with SN 1996al, the spectra of SN 2013gc show an increasing strength of the He lines from the early to the late phases, which may indicate the presence of higher-density clumps heated by the SN shock in its lower-density, spherically symmetric CSM component.
We also note that the radial velocities of the H$\alpha$ bumps of SN 1996al and SN 2013gc are comparable, supporting an overall similarity in the gas ejection scenarios for the two objects.
SN 1996al was characterized by low ejecta mass and a modest kinetic energy. The ejecta interaction with a dense CSM embedding the progenitor determined the light curve features at all phases, and was still active 15 years after the explosion. The proposed CSM structure around SN 2013gc is composed by an equatorial disc with a complex density profile, and ejecta-CSM interaction being always present. When the ejecta reach the first CSM equatorial density enhancement, the conversion of kinetic energy into radiation gives rise to the observed plateau, while the encounter with an outer layer with higher density powers the second peak. Later on, ejecta-CSM interactions weakens and the SN light curve finally starts the fast decline.

\section{Conclusions}
\label{conclusion}
SN 2013gc is a member of the type IId SNe class, a subgroup of type IIn SNe whose spectra are characterized by double, broad and narrow, P-Cygni H$\alpha$ components. SN 2013gc can be considered a scaled-down version of SN 1996al, triggered by the same physical process: a fall-back SN from a highly massive star, possibly an LBV. SN 2013gc is the first object of this class showing a long-duration progenitor's eruptive phase lasting a few years (at least, from 2010 and 2013), and continued until a very short time before the SN explosion. The object showed significant variability, with oscillations of 1-2 mag in very short time-scales (days to months). The multiple flares and the long-duration major eruption are best suited with a massive progenitor in its final evolutionary stages, very likely an LBV. 
The outbursts formed a structured CSM around the progenitor (and very close to it), and the SN ejecta interact with it from soon after the SN explosion. When the ejecta collide with denser CSM layers, the interaction powered the observed light curve plateau and the second peak. The CSM is likely clumpy, as deduced from the evolution of the He lines, and is composed of a spherically symmetric component, two denser shells and a further bipolar CSM component. The fall-back scenario of a massive star is supported by the low $^{56}$Ni mass found, the low luminosity of the SN, the lack of [\ion{O}{I}] lines in the spectra and the detection of numerous pre-SN outbursts from the progenitor.
We note, however, thst the observables of SN 2013gc are not inconsistent with a non-terminal outburst, followed by shell-shell collision.

\section*{Agknowledgment}
\begin{small}
We thank the anonymous referee for helpful comments, that improved our paper.

This research has made use of the NASA/IPAC Extragalactic Database (NED) which is operated by the Jet Propulsion Laboratory, California Institute of Technology, under contract with the National Aeronautics and Space Administration.

Support for G.P. and F.O.E. is provided by the Ministry of Economy, Development, and Tourism's Millennium Science Initiative through grant IC120009, awarded to The Millennium Institute of Astrophysics (MAS). F.O.E acknowledges support from the FONDECYT grant N$^{\circ}$ 11170953.
The CHASE project is founded by the Millennium Institute for Astrophysics.

Based in part on data obtained from the ESO Science Archive Facility under program IDs 188.D-3003, 177.D-3023 for SN 2013gc, 65.H-0292(D), 66.D-0683(C), 69.D-0672(A), 67.D-0422(B), 71.D-0265(A), 65.I-0319(A), 65.N-0287(A), 67.D-0438(B), 77.B-0741(A) for SN 2000P. We thank G. Altavilla for the support on the observations of SN 2000P.

Based in part on observations obtained at the Southern Astrophysical Research (SOAR) telescope, which is a joint project of the Minist\'{e}rio da Ci\^{e}ncia, Tecnologia, Inova\c{c}\~{a}os e Comunica\c{c}\~{a}oes (MCTIC) do Brasil, the U.S. National Optical Astronomy Observatory (NOAO), the University of North Carolina at Chapel Hill (UNC), and Michigan State University (MSU).

Based in part on observations at Cerro Tololo Inter-American Observatory, National Optical Astronomy Observatory (NOAO), which is operated by the Association of Universities for Research in Astronomy (AURA), Inc. under a cooperative agreement with the National Science Foundation. 
The Dark Energy Camera Plane Survey (DECaPS; NOAO Program Number 2016A-0323 and 2016B-0279, PI: Finkbeiner) includes data obtained at the Blanco telescope, Cerro Tololo Inter-American Observatory, National Optical Astronomy Observatory (NOAO). 

The SARA Observatory is supported by the National Science Foundation (AST-9423922), the Research Corporation and the State of Florida Technological Research and Development Authority.

This research has made use of the NASA/IPAC Infrared Science Archive, which is operated by the Jet Propulsion Laboratory, California Institute of Technology, under contract with the National Aeronautics and Space Administration.

Based in part on observations acquired through the Gemini Observatory Archive.

The Pan-STARRS1 Surveys (PS1) and the PS1 public science archive have been made possible through contributions by the Institute for Astronomy, the University of Hawaii, the Pan-STARRS Project Office, the Max-Planck Society and its participating institutes, the Max Planck Institute for Astronomy, Heidelberg and the Max Planck Institute for Extraterrestrial Physics, Garching, The Johns Hopkins University, Durham University, the University of Edinburgh, the Queen's University Belfast, the Harvard-Smithsonian Center for Astrophysics, the Las Cumbres Observatory Global Telescope Network Incorporated, the National Central University of Taiwan, the Space Telescope Science Institute, the National Aeronautics and Space Administration under Grant No. NNX08AR22G issued through the Planetary Science Division of the NASA Science Mission Directorate, the National Science Foundation Grant No. AST-1238877, the University of Maryland, Eotvos Lorand University (ELTE), the Los Alamos National Laboratory, and the Gordon and Betty Moore Foundation.

IRAF is distributed by the National Optical Astronomy Observatory, which is operated by the AURA.
This research has made use of NASA's Astrophysics Data System Bibliographic Services.
\end{small}

\appendix
\section{}
\label{2000p}
\subsection{Light curves and spectroscopic sequence of SN 2000P}
SN 2000P was discovered on 2000 March 8 by the amateur astronomer R. Chassagne at R.A.=$13^h$07$^m$9$^s$.88 and Dec=$-28^{\circ}$13'59".3 (J2000). The SN was located about 16" east and 21" south of the center of the spiral galaxy NGC 4965 \citep{2000IAUC.7378....1C}. The confirmation image was taken at the Pic du Midi Observatory.
The discovery magnitude was 14.1.
The original classification as type IIn supernova was obtained through an optical spectrum taken on 2000 March 10 with the ESO La Silla 1.54-m Danish telescope+DFOSC \citep{2000IAUC.7380....2C}.
The redshift of the galaxy and the mean distance modulus, as reported in NED, are $z=0.007542\pm0.000017$ and $\mu=32.83\pm0.77$ mag, respectively.

\begin{figure}
\includegraphics[width=1.1\columnwidth]{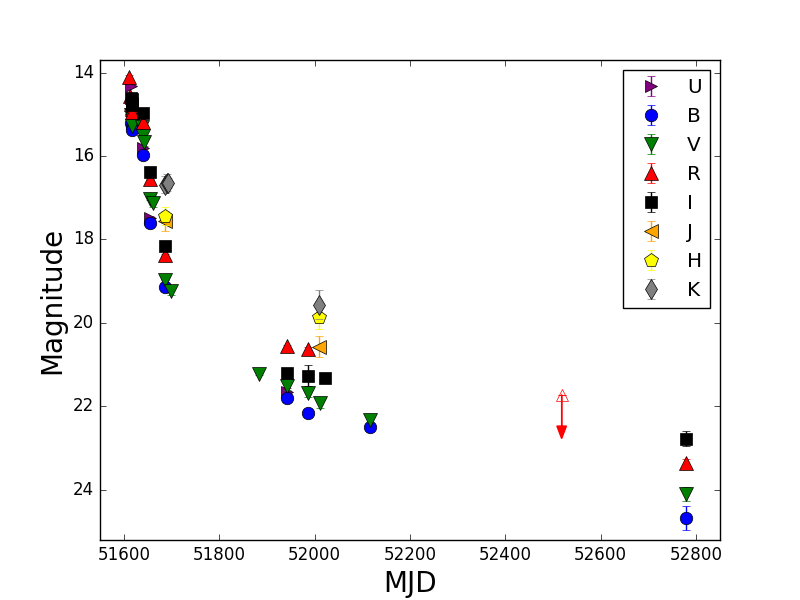}
\caption{Optical and NIR light curves of SN 2000P, spanning about 3 years of observations. The magnitudes are not corrected for line-of-sight extinction. The red arrow indicates an upper limit in the $R$ band.}
\label{2000P}
\end{figure}

\begin{figure*}
\includegraphics[width=2\columnwidth]{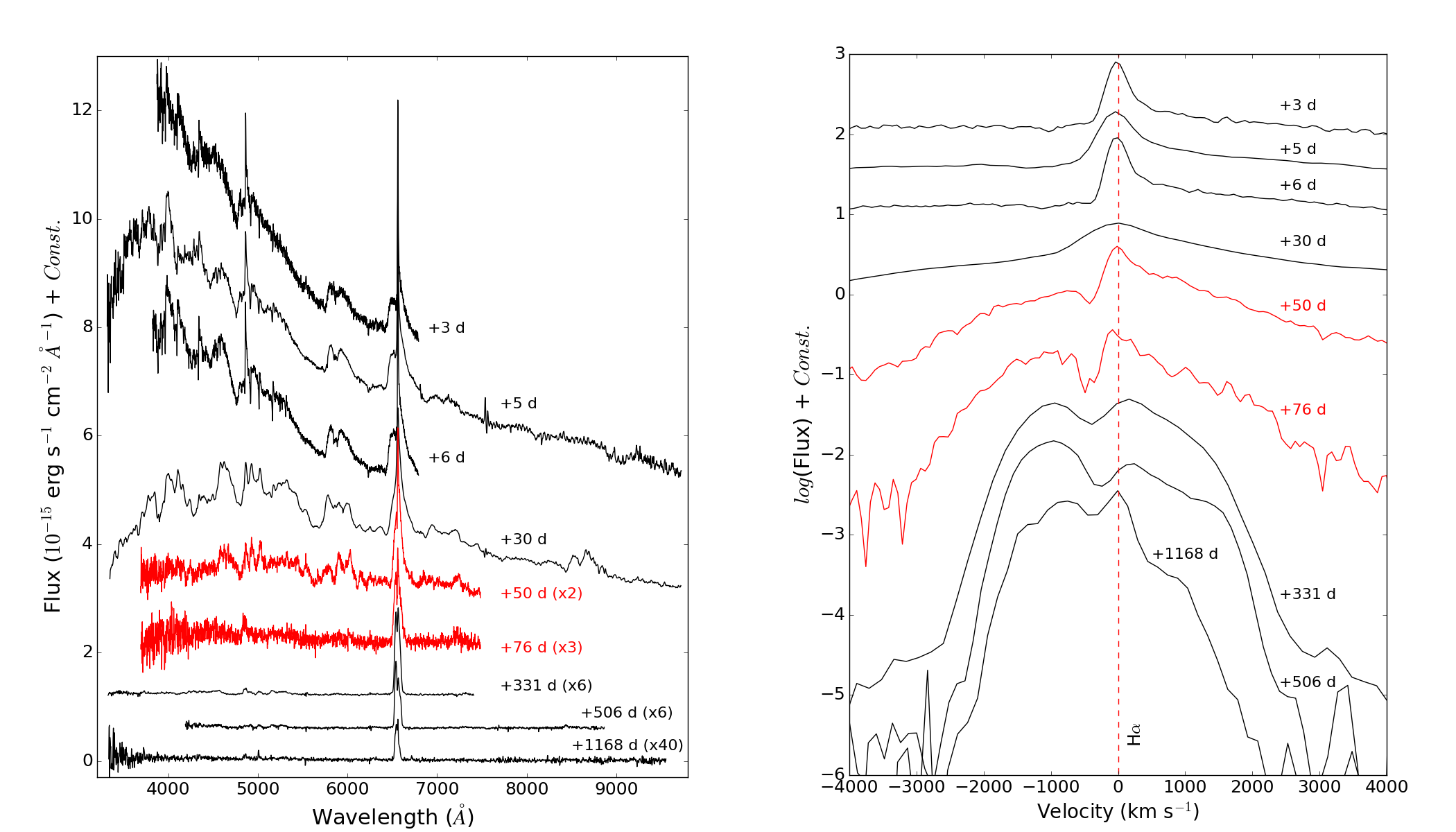}
\caption{Left: Time sequence of optical spectra of SN 2000P, spanning about 3 years of observations. The fluxes of the spectra are scaled to the $R$-band or, if not available, to the $V$-band photometry of the nearest night. The spectra plotted in red colour are taken from CfA Supernova Data Archive (https://www.cfa.harvard.edu/supernova/SNarchive.html), to fill the time gap of our sequence. Right: Evolution of the H$\alpha$ profile in the same spectral sequence. The logarithm of the specific flux is reported on the $y$-axis. The velocity with respect to the rest-frame is shown. In both graphs, the phase is relative to the discovery.}
\label{seq}
\end{figure*}

The \textit{UBVRIJHK} photometry of SN 2000P is presented in Tab. \ref{tab44}, the technical informations on the spectra are provided in Tab. \ref{tab55}. In Fig. \ref{2000P} we give the optical and NIR light curve of SN 2000P, while in Fig. \ref{seq} we plot the spectral sequence and the evolution of the H$\alpha$ profile.

\begin{landscape}
\begin{table}
\caption{Optical and NIR photometric measurements for SN 2000P. For completeness, amateur astronomers measurements are also included.}
\label{tab44}
\begin{threeparttable}[b]
\begin{tabular}{lccccccccccll}
\hline
Date & MJD & phase\tnote{a} & \emph{U} & \emph{B} & \emph{V} & \emph{R} & \emph{I} & \emph{J} & \emph{H} & \emph{K} & Telescope\tnote{b} & Observation PI\\
\hline
2000-03-08 & 51611.03 & 0.0 & & & & 14.10(04) & & & & & 0.3 meter & Chassagne\\
2000-03-09 & 51612.12 & 1.1 & & & & 14.57(05) & & & & & Pic & Colas\\
2000-03-10 & 51614.18 & 3.2 & 14.31(03) & 15.20(02) & 15.01(02) & 14.76(02) & 14.61(02) & & & & Dan & Turatto\\
2000-03-12 & 51616.24 & 5.2 & 14.68(03) & 15.30(02) & 15.10(02) & 14.86(02) & 14.64(02) & & & & Dan & Turatto\\
2000-03-13 & 51616.64 & 5.6 & & & 15.17(05) & & 14.79(05) & & & & 0.25 meter  & Kiyota\\
2000-03-13 & 51617.24 & 6.2 & 14.91(03) & 15.37(02) & 15.29(02) & 14.94(02) & 14.73(02) & & & & Dan & Turatto\\
2000-04-04 & 51638.01 & 27.0 & & & 15.30(10) & & & & & & KAIT & \citet{2002PASP..114..403L}\\
2000-04-07 & 51641.08 & 30.1 & 15.82(05) & 15.98(05) & 15.52(02) & 15.20(03) & 14.98(07) & & & & EF2 & Pastorello\\
2000-04-09 & 51643.04 & 32.0 & & & 15.67(10) & & & & & & KAIT & \citet{2002PASP..114..403L}\\
2000-04-19 & 51654.00 & 43.0 & 17.48(03) & 17.60(03) & 17.04(03) & 16.56(02) & 16.39(02) & & & & TNG & Benetti\\
2000-04-26 & 51661.25 & 50.2 & & & 17.12(10) & & & & & & KAIT & \citet{2002PASP..114..403L}\\
2000-05-23 & 51687.00 & 76.0 & & 19.14(05) & 18.96(02) & 18.37(02) & 18.17(05) & & & & TNG & Benetti\\
2000-05-23 & 51687.09 & 76.1 & & & & & & 17.55(24) & 17.44(21) & 16.69(20) & SOFI & Salamanca\\
2000-05-26 & 51690.95 & 79.9 & & & & & & & & 16.64(20) & SOFI & Grosbol\\
2000-05-27 & 51691.95 & 79.9 & & & & & & & & 16.65(12) & SOFI & Grosbol\\
2000-06-05 & 51699.50 & 88.5 & & & 19.23(10) & & & & & & KAIT & \citet{2002PASP..114..403L}\\
2000-12-06 & 51884.60 & 273.6 & & & 21.22(02)\tnote{c} & & & & & & HST & \citet{2002PASP..114..403L}\\
2001-02-02 & 51942.37 & 331.3 & 21.65(08) & 21.80(09) & 21.51(08) & 20.56(03) & 21.20(06) & & & & EF2 & Pastorello\\
2001-03-17 & 51985.35 & 374.3 & & 22.17(10) & 21.69(08) & 20.62(04) & & & & & Dan & Pastorello\\
2001-03-18 & 51986.29 & 375.2 & & & & & 21.26(26) & & & & Dan & Pastorello\\
2001-04-09 & 52008.30 & 397.3 & & & & & & 20.57(25) & 19.85(30) & 19.56(34) & SOFI & Spyromilio\\
2001-04-12 & 52012.15 & 401.1 & & & 21.92(12) & & & & & & Dan & Pastorello\\
2001-04-23 & 52022.10 & 411.1 & & & & & 21.33(03)\tnote{d} & & & & HST & \citet{2002PASP..114..403L}\\
2001-07-26 & 52117.02 & 506.0 & & 22.49(04) & 22.33(05) & & & & & & VLT1 & Cappellaro\\
2002-08-31 & 52518.15 & 907.1 & & & & >21.72\tnote{e} & & & & & EF2 & Pastorello\\
2003-05-20 & 52779.02 & 1168.0 & & 24.67(29) & 24.10(16) & 23.36(11) & 22.78(18) & & & & VLT2 & Zampieri\\
\hline
\end{tabular}
\begin{tablenotes}
\item[a] Days from the discovery \item[b] Pic = Gentili 1.05m Pic du Midi, Dan = ESO Danish 1.54m+DFOSC, EF2 = ESO 3.6m+EFOSC2, TNG = TNG 3.6m+OIG, SOFI = ESO NTT+SOFI, VLT1 = VLT (UT1) 8.2m+FORS1, VLT2 = VLT (UT4) 8.2m+FORS2
\item[c] HST+F555W \item[d] HST+F814W \item[e] Upper limit
\end{tablenotes}
\end{threeparttable}
\end{table}
\end{landscape}

\begin{table*}
\caption{Spectroscopic observations of SN 2000P.}
\label{tab55}
\begin{threeparttable}[b]
\begin{tabular}{lcclccccll}
\hline
Date & & MJD & & phase\tnote{a} & & Range & Resolution\tnote{b} & Telescope & Instrument\\
 & & & & (d) & & (\AA) & (\AA) & &\\
\hline
2000-03-11 & & 51614.2 & & 3.2   & & 3900-6840 & 4.7 & ESO 1.54m & DFOSC (gr7)\\
2000-03-13 & & 51616.3 & & 5.3   & & 3350-9800 & 9.3 & ESO 1.54m & DFOSC (gr4+gr5)\\
2000-03-14 & & 51617.2 & & 6.2   & & 3860-6820 & 5.1 & ESO 1.54m & DFOSC (gr7)\\
2000-04-07 & & 51641.1 & & 29.1  & & 3370-9800 & 16 & ESO 3.6m & EFOSC2 (gr11+gr12)\\
2001-02-02 & & 51942.3 & & 331.3 & & 3360-7450 & 9  & ESO 3.6m & EFOSC2 (gr11)\\
2001-07-27 & & 52117.0 & & 506.0 & & 4230-8920 & 11 & VLT UT1 8.2m & FORS1\\
2003-05-20 & & 52779.0 & & 1168.0 & & 3400-9600 & 10 & VLT UT4 8.2m & FORS2\\
\hline
\end{tabular}
\begin{tablenotes}
\item[a] Days from the discovery \item[a] FWHM of the night sky lines.
\end{tablenotes}
\end{threeparttable}
\end{table*}

\begin{table*}
\caption{Optical Johnson-Cousins and Sloan photometry of SN 2013gc.}
\label{tab5}
\begin{threeparttable}[b]
\begin{tabular}{llcccccccl}
\hline
Date & MJD & \emph{B} & \emph{V} & \emph{R} & \emph{I} & \emph{g} & \emph{z} & \emph{y} & Telescope\tnote{a}\\
\hline
2010-03-02 & 55257.32 & & & & & & & 19.57(0.37) & PS1 \\
2010-04-04 & 55290.24 & & & & & & 20.32(0.15) & & PS1 \\
2010-11-02 & 55517.61 & & & & & & >20.11 & & PS1 \\
2010-11-17 & 55517.61 & & & & & & & >18.90 & PS1 \\
2010-12-24 & 55554$\pm22$ & & & 21.78(0.40) & & & & & PTF \\
2011-01-18 & 55579$\pm1$ & & & >20.42 & & & & & PTF \\
2011-01-24 & 55585$\pm3$ & & & >18.96 & & & & & PTF \\
2011-03-25 & 55645.24 & & & & & & & 19.47(0.30) & PS1 \\
2011-10-20 & 55854.65 & & & & & & & 19.06(0.16) & PS1 \\
2011-11-14 & 55879.60 & & & & & & 21.06(0.29) & & PS1 \\
2012-02-10 & 55967.37 & & & & 19.76(0.08)\tnote{d} & & & & PS1 \\
2012-04-04 & 56021.24 & & & & & & & >19.72 & PS1 \\
2012-10-29 & 56229.63 & & & & & & 20.64(0.31) & & PS1 \\
2012-10-30 & 56230.65 & & & & & & & >19.62 & PS1 \\
2012-11-14\tnote{e} & 56245.36 & & & 20.89(0.16)\tnote{b} & 20.99(0.25)\tnote{b} & & & & VST \\
2012-12-28 & 56289.54 & & & & 20.19(0.18)\tnote{d} & & & & PS1 \\
2013-03-11 & 56362.28 & & >18.83 & & & & & & NTT \\
2013-03-19 & 56370.16 & & >19.39 & & & & & & NTT \\
2013-03-29 & 56380.98 & >18.55 & >18.65 & >18.44 & >18.31 & & & & PROMPT \\
2013-04-01 & 56383.18 & >18.37 & & & & & & & PROMPT \\
2013-04-01 & 56383.98 & >19.21 & >18.56 & >18.64 & >18.13 & & & & PROMPT \\
2013-04-02 & 56384.14 & & >20.33 & & & & & & NTT \\
2013-04-02 & 56384.98 & >18.30 & >18.52 & >18.13 & >17.91 & & & & PROMPT \\
2013-04-04 & 56386.97 & >19.00 & >18.76 & >18.52 & >18.26 & & & & PROMPT \\
2013-04-05 & 56387.98 & >19.95 & >18.82 & & >17.91 & & & & PROMPT \\
2013-04-06 & 56388.08 & & >20.58 & & & & & & NTT \\
2013-04-07 & 56389.01 & >20.01 & >19.61 & & & & & & PROMPT \\
2013-04-08 & 56390.00 & >20.22 & >18.84 & >19.22 & >18.09 & & & & PROMPT \\
2013-04-08 & 56390.97 & & >17.58 & >17.79 & >17.66 & & & & PROMPT \\
2013-04-10 & 56392.98 & >19.71 & >18.85 & >18.82 & >18.26 & & & & PROMPT \\
2013-04-11 & 56393.97 & >18.33 & & & & & & & PROMPT \\
2013-04-12 & 56394.98 & >20.28 & & & & & & & PROMPT \\
2013-04-13 & 56395.05 & & >20.35 & & & & & & NTT \\
2013-04-14 & 56396.03 & & & & & >22.37 & & & VST \\
2013-04-14 & 56396.04 & >18.87 & >18.97 & >19.28 & >18.34 & & & & PROMPT \\
2013-04-14 & 56396.96 & >18.81 & >18.00 & >18.27 & & & & & PROMPT \\
2013-04-16 & 56398.96 & >19.25 & >19.56 & >19.35 & >18.78 & & & & PROMPT \\
2013-04-19 & 56401.02 & & >20.29 & & & & & & NTT \\
2013-04-19 & 56401.98 & >20.99 & >20.52 & >20.24 & >19.72 & & & & PROMPT \\
2013-04-20 & 56402.96 & >20.42 & >20.59 & >20.58 & >19.98 & & & & PROMPT \\
2013-04-22 & 56404.05 & >20.87 & >19.09 & >19.15 & >18.60 & & & & PROMPT \\
2013-04-25 & 56407.98 & >19.42 & >19.33 & >19.67 & >19.25 & & & & PROMPT \\
2013-04-28 & 56410.08 & & >17.64 & & & & & & PROMPT \\
2013-04-28 & 56410.96 & & >17.99 & >18.34 & & & & & PROMPT \\
2013-04-29 & 56411.96 & & >18.70 & & & & & & PROMPT \\
2013-05-01 & 56413.95 & >19.30 & & & & & & & PROMPT \\
2013-05-04 & 56416.06 & >19.64 & >18.20 & & & & & & PROMPT \\
2013-05-04 & 56416.96 & & & 20.45(0.14) & & & & & GEMINI \\
2013-05-05 & 56417.99 & >20.74 & >18.84 & & & & & & PROMPT \\
2013-05-07 & 56419.01 & >20.48 & >19.16 & >19.07 & & & & & PROMPT \\
2013-05-11 & 56423.96 & >20.66 & >21.07 & >20.43 & >20.29 & & & & PROMPT \\
2013-05-15 & 56427.96 & >19.80 & >19.99 & & & & & & PROMPT \\
2013-05-19 & 56431.96 & >19.99 & >20.23 & >18.71 & & & & & PROMPT \\
2013-05-20 & 56432.97 & & & >20.13 & >19.67 & & & & PROMPT \\
2013-05-22 & 56434.97 & >20.90 & & & & & & & PROMPT \\
2013-05-23 & 56435.96 & >19.44 & >20.99 & & & & & & PROMPT \\
2013-05-24 & 56436.96 & & & >20.12 & >19.30 & & & & PROMPT \\
2013-05-29 & 56441.96 & >18.96 & >20.13 & >20.75 & >19.17 & & & & PROMPT \\
2013-06-03 & 56446.94 & & >18.15 & & & & & & PROMPT \\
2013-06-12 & 56455.95 & & >19.06 & >19.79 & & & & & PROMPT \\
2013-06-19 & 56462.94 & >18.23 & >18.66 & & & & & & PROMPT \\
2013-08-26 & 56530.39 & & & 18.21(0.08) & & & & & SOAR \\
2013-09-12 & 56547.36 & & 15.16(0.02) & & & & & & NTT \\ 
2013-09-13 & 56548.36 & & 15.14(0.04) & & & & & & NTT \\
\hline
\end{tabular}
\end{threeparttable}
\end{table*}
\begin{table*}
\contcaption{Optical Johnson and Sloan photometry of SN 2013gc.}
\begin{threeparttable}[b]
\begin{tabular}{llcccccccl}
\hline
Date & MJD & \emph{B} & \emph{V} & \emph{R} & \emph{I} & \emph{g} & \emph{z} & \emph{y} & Telescope\tnote{a}\\
\hline
2013-09-28 & 56563.38 & & & & & 15.95(0.05) & & & PROMPT \\
2013-09-29 & 56564.40 & & & & 14.66(0.14)\tnote{b} & 15.91(0.06) & & & PROMPT \\
2013-10-08 & 56573.29 & & 16.59(0.05) & & & & & & NTT \\
2013-11-04 & 56600.65 & & & & & & 16.71(0.02) & 16.60(0.03) & PS1 \\
2013-11-06 & 56602.28 & 19.93(0.04) & 18.75(0.08) & & & & & & SARA \\
2013-11-08 & 56604.30 & & & 17.67(0.11)\tnote{b} & 17.18(0.08)\tnote{b} & 19.54(0.08) & 16.83(0.06) & & PROMPT \\
2013-11-12 & 56608.31 & & & 17.84(0.07) & 17.41(0.10) & & & & TRAP \\
2013-11-16 & 56612.26 & & & 18.29(0.05) & 17.53(0.10) & & & & SARA \\ 
2013-11-18 & 56614.25 & & & 18.40(0.06) & & & & & SOAR \\
2013-11-26 & 56622.28 & 20.86(0.37) & 20.04(0.13) & & & & & & SARA \\
2013-12-29 & 56655.17 & & & 19.01(0.06) & 18.41(0.11) & & & & SARA \\
2014-01-02 & 56659.22 & 21.26(0.12) & 20.30(0.04) & & & & & & TRAP \\ 
2014-01-12 & 56669.19 & & & 19.02(0.24) & 18.44(0.21) & & & & TRAP \\
2014-02-02 & 56690.17 & & & 18.96(0.22) & 18.49(0.12) & & & & TRAP \\
2014-02-13 & 56701.15 & & 20.22(0.19) & 18.85(0.05) & & & & & TRAP \\
2014-03-02 & 56718.11 & & & 18.83(0.07) & 18.52(0.07) & & & & TRAP \\
2014-03-22 & 56738.05 & & & & 17.38(0.04) & & & & TRAP \\
2014-04-02 & 56749.00 & & 19.91(0.09) & & & & & & TRAP \\
2014-05-04 & 56781.24 & & & & & & & 18.35(0.07) & PS1 \\
2014-05-05 & 56782.09 & & 19.79(0.11) & & & & & & PONT \\
2014-05-08 & 56785.97 & 20.59(0.32) & 20.18(0.27) & 18.64(0.13) & 18.41(0.09) & & & & SARA \\
2014-05-09 & 56786.99 & & & 18.66(0.15)\tnote{b} & 18.43(0.30)\tnote{b} & 20.59(0.33) & 17.72(0.30) & & PROMPT \\
2014-05-11 & 56788.98 & >20.52 & 20.09(0.14) & 18.68(0.14) & 18.38(0.09) & & & & TRAP \\
2014-05-11 & 56788.99 & & & 18.77(0.06) & & & & & SOAR \\
2014-05-16 & 56793.98 & 20.82(0.27) & 20.13(0.14) & & & 20.58(0.23) & & & SARA \\
2014-05-27 & 56804.96 & 20.54(0.33) & 19.93(0.20) & 19.04(0.13) & 18.52(0.10) & & & & SARA \\
2014-05-30 & 56807.98 & 20.91(0.16) & 19.83(0.10) & 18.56(0.08) & & & & & SMARTS \\
2014-06-01 & 56809.97 & 20.88(0.18) & 20.01(0.13) & 18.67(0.11) & 18.18(0.20) & & & & TRAP \\
2014-06-10 & 56818.97 & >20.93 & & & & & & & TRAP \\
2014-06-26 & 56834.97 & & & 19.28(0.19) & 18.84(0.23) & & & & SARA \\
2015-02-22 & 57075.05 & >21.01 & >21.00 & & >19.46 & & & & TRAP \\
2015-04-02 & 57114.02 & >21.21 & >20.94 & 20.84(0.46) & >20.34 & & & & TRAP \\
2015-05-01 & 57143.99 & >20.45 & >20.54 & 20.95(0.32) & >19.93 & & & & TRAP \\
2015-05-12 & 57154.00 & >21.26 & & & & & & & TRAP \\
2015-05-20 & 57163.00 & & & 21.04(0.18) & >20.13 & & & & TRAP \\
2015-06-01 & 57174.99 & >20.37 & >20.26 & 21.18(0.37) & >20.12 & & & & TRAP \\
2015-06-11 & 57185.00 & >20.78 & & & & & & & TRAP \\
2017-03-04\tnote{c} & 57816$\pm47$ & & & & & 23.57(0.22) & 22.62(0.24) & & DEC \\
\hline
\end{tabular}
\begin{tablenotes}
\item[a] NTT = ESO 3.6-meter NTT+EFOSC2, SOAR = 4.1-meter `SOAR'+Goodman Spectrograph, TRAP = 0.5-meter TRAPPIST, SMARTS = CTIO 1.3-meter+ANDICAM, SARA = CTIO 0.6-meter+ARC, GEMINI = 8.1-meter Gemini South+GMOS-S, PTF = 1.2-meter `S. Oschin' Schmidt+PTF survey, PONT = Las Campanas 2.5-meter `Du Pont'+WFCCD/WF4K-1, PS1 = Haleakala 1.8-meter+Pan-STARRS1 Survey, DEC = CTIO 4-meter 'V. Blanco'+Dark Energy Camera (DECaPS survey).
\item[b] Converted from Sloan to Johnson photometric system. \item[c] For this epoch we also report $r$ = 22.63(0.18). \item[d] PS1 $i$ band magnitude, not converted to $I$. \item[e] For this epoch and this instrument we also report $u$ > 21.40.
\end{tablenotes} 
\end{threeparttable}
\end{table*}

\begin{table*}
\caption{NIR photometry of SN 2013gc.}
\label{tab7}
\begin{threeparttable}[b]
\begin{tabular}{llcccl}
\hline
Date & MJD & \emph{J} & \emph{H} & \emph{K} & Telescope\tnote{a}\\
\hline
2013-03-18 & 56370.00 & >19.70 & >18.02 & >18.32 & SOFI \\
2013-04-04 & 56386.02 & 19.29(0.33) & 18.84(0.32) & 18.08(0.19) & SOFI \\
2013-04-12 & 56394.05 & 19.39(0.26) & 18.74(0.40) & 17.91(0.25) & SOFI \\
2013-04-18 & 56400.05 & 19.46(0.27) & 18.81(0.33) & 18.54(0.37) & SOFI \\
2014-05-30 & 56807.98 & 17.29(0.16) & 17.15(0.18) & 17.55(0.30) & SMARTS \\
\hline
\end{tabular}
\begin{tablenotes}
\item[a] SOFI = ESO 3.6-meter NTT+SOFI, SMARTS = CTIO 1.3-meter+ANDICAM.
\end{tablenotes}
\end{threeparttable}
\end{table*}

\begin{table}
\caption{Clear photometry of SN 2013gc, treated as $R$-band.}
\label{tab6}
\begin{tabular}{lccl}
Date & MJD & clear & Telescope \\
\hline
2011-01-27 & 55588.07 &   >18.99    & PROMPT \\
2011-02-01 & 55593.08 &   >19.32    & PROMPT \\
2011-02-20 & 55612.10 &   >19.72    & PROMPT \\
2011-03-06 & 55626.06 & 19.70(0.15) & PROMPT \\
2011-03-12 & 55632.03 &   >19.68    & PROMPT \\
2011-05-02 & 55683.08 &   >19.57    & PROMPT \\
2011-09-14 & 55818.37 &   >19.01    & PROMPT \\
2011-09-25 & 55829.34 &   >19.57    & PROMPT \\
2011-10-17 & 55851.28 &   >19.08    & PROMPT \\
2011-10-22 & 55856.35 &   >18.49    & PROMPT \\
2011-11-02 & 55867.24 &   >19.47    & PROMPT \\
2012-01-04 & 55930.10 &   >19.16    & PROMPT \\
2012-01-07 & 55933.19 & 20.36(0.36) & PROMPT \\
2012-01-10 & 55936.20 &   >19.27    & PROMPT \\
2012-01-14 & 55940.21 & 20.42(0.27) & PROMPT \\
2012-01-18 & 55944.18 &   >20.01    & PROMPT \\
2012-01-24 & 55950.15 &   >19.95    & PROMPT \\
2012-01-27 & 55953.15 & 20.46(0.41) & PROMPT \\
2012-01-29 & 55955.13 &   >19.72    & PROMPT \\
2012-02-01 & 55958.20 & 20.93(0.41) & PROMPT \\
2012-02-03 & 55960.16 &   >19.48    & PROMPT \\
2012-02-06 & 55963.12 &   >19.57    & PROMPT \\
2012-02-08 & 55965.14 & 19.89(0.45) & PROMPT \\
2012-02-10 & 55967.11 & 19.93(0.37) & PROMPT \\
2012-02-12 & 55969.10 & 20.35(0.34) & PROMPT \\
2012-02-21 & 55978.10 &   >20.16    & PROMPT \\
2012-02-24 & 55981.07 &   >19.39    & PROMPT \\
2012-02-25 & 55982.09 & 21.09(0.50) & PROMPT \\
2012-02-26 & 55983.08 & 20.00(0.20) & PROMPT \\
2012-02-27 & 55984.09 & 20.65(0.26) & PROMPT \\
2012-03-02 & 55988.09 &   >19.64    & PROMPT \\
2012-03-05 & 55991.07 &   >19.90    & PROMPT \\
2012-03-07 & 55993.05 &   >19.19    & PROMPT \\
2012-03-09 & 55995.07 & 19.96(0.37) & PROMPT \\
2012-03-11 & 55997.06 &   >19.26    & PROMPT \\
2012-03-13 & 55999.05 &   >19.95    & PROMPT \\
2012-03-14 & 56000.06 &   >18.70    & PROMPT \\
2012-03-16 & 56002.05 & 21.27(0.46) & PROMPT \\
2012-03-18 & 56004.04 &   >20.14    & PROMPT \\
2012-03-26 & 56012.03 &   >20.12    & PROMPT \\
2012-03-31 & 56017.01 & 20.29(0.47) & PROMPT \\
2012-04-06 & 56023.01 &   >19.63    & PROMPT \\
2012-10-11 & 56211.30 &   >19.67    & PROMPT \\
2012-10-14 & 56214.29 &   >19.71    & PROMPT \\
2012-11-27 & 56258.22 &   >19.72    & PROMPT \\
2012-12-01 & 56262.25 & 19.93(0.32) & PROMPT \\
2012-12-04 & 56265.30 &   >19.43    & PROMPT \\
2012-12-06 & 56267.29 &   >19.75    & PROMPT \\ 
2012-12-09 & 56270.15 &   >19.09    & PROMPT \\
2012-12-13 & 56274.20 &   >19.86    & PROMPT \\
2012-12-16 & 56277.12 &   >19.80    & PROMPT \\
2012-12-21 & 56282.12 & 20.47(0.42) & PROMPT \\ 
2012-12-24 & 56285.16 & 20.41(0.39) & PROMPT \\
2012-12-27 & 56288.23 & 20.10(0.44) & PROMPT \\
2012-12-31 & 56292.22 &   >19.69    & PROMPT \\
2013-01-04 & 56296.21 &   >19.48    & PROMPT \\
2013-01-13 & 56305.22 &   >19.62    & PROMPT \\
2013-01-16 & 56308.12 &   >19.69    & PROMPT \\
2013-01-23 & 56315.35 &   >19.36    & PROMPT \\
2013-01-26 & 56318.14 &   >19.37    & PROMPT \\
2013-02-02 & 56325.21 &   >19.56    & PROMPT \\
2013-02-08 & 56331.30 &   >19.62    & PROMPT \\
2013-02-15 & 56338.15 &   >20.02    & PROMPT \\
2013-03-09 & 56360.23 &   >19.79    & PROMPT \\
2013-03-10 & 56361.12 &   >19.50    & PROMPT \\
2013-03-11 & 56362.14 &   >19.01    & PROMPT \\
2013-03-17 & 56368.08 &   >19.71    & PROMPT \\
2013-03-19 & 56370.22 &   >19.52    & PROMPT \\
2013-03-23 & 56374.18 &   >19.21    & PROMPT \\
2013-03-26 & 56377.18 &   >19.25    & PROMPT \\
\hline
\end{tabular}
\end{table}
\begin{table}
\contcaption{Clear photometry of SN 2013gc.}
\begin{tabular}{lccl}
Date & MJD & clear & Telescope \\
\hline
2013-03-28 & 56379.12 &   >19.58    & PROMPT \\
2013-04-04 & 56386.07 &   >19.51    & PROMPT \\
2013-04-21 & 56403.03 &   >19.69    & PROMPT \\
2013-11-07 & 56603.33 & 17.59(0.08) & PROMPT \\
2013-11-08 & 56604.25 & 17.78(0.07) & PROMPT \\
2013-11-09 & 56605.23 & 17.77(0.08) & PROMPT \\
2013-11-11 & 56607.22 & 17.83(0.06) & PROMPT \\
2013-11-14 & 56610.21 & 18.15(0.14) & PROMPT \\
2013-11-17 & 56613.27 & 18.31(0.15) & PROMPT \\
2013-11-19 & 56615.24 & 18.52(0.13) & PROMPT \\
2013-11-21 & 56617.19 & 19.06(0.15) & PROMPT \\
2013-11-24 & 56620.31 & 19.10(0.18) & PROMPT \\
2013-11-26 & 56622.28 & 19.08(0.13) & PROMPT \\
2013-11-30 & 56626.20 & 19.14(0.18) & PROMPT \\
2013-12-02 & 56628.16 & 19.16(0.17) & PROMPT \\
2013-12-06 & 56632.15 & 19.19(0.20) & PROMPT \\
2013-12-12 & 56638.14 & 19.29(0.13) & PROMPT \\ 
2014-03-12 & 56728.21 & 18.21(0.11) & PROMPT \\ 
2014-04-10 & 56757.09 & 18.13(0.07) & PROMPT \\
2014-04-11 & 56758.14 & 18.34(0.10) & PROMPT \\
2014-04-16 & 56763.08 & 18.50(0.19) & PROMPT \\
2014-04-17 & 56764.08 & 18.70(0.15) & PROMPT \\
2014-04-19 & 56766.06 & 18.61(0.08) & PROMPT \\
2014-04-20 & 56767.09 & 18.77(0.11) & PROMPT \\
2014-04-22 & 56769.07 & 18.64(0.06) & PROMPT \\
2014-05-13 & 56790.05 & 19.14(0.17) & PROMPT \\
2014-05-16 & 56793.00 & 19.07(0.10) & PROMPT \\
2014-09-07 & 56907.40 & 19.99(0.38) & PROMPT \\
2014-09-24 & 56924.36 & 20.05(0.37) & PROMPT \\
2014-10-03 & 56933.37 & 20.11(0.41) & PROMPT \\
2014-10-09 & 56939.31 &   >18.21    & PROMPT \\
2014-10-11 & 56941.30 &   >18.47    & PROMPT \\
2014-10-20 & 56950.28 & 20.28(0.33) & PROMPT \\
2014-11-08 & 56969.22 &   >18.99    & PROMPT \\
2014-11-13 & 56974.32 &   >19.80    & PROMPT \\
2015-01-03 & 57025.20 &   >19.54    & PROMPT \\ 
2015-01-07 & 57029.19 &   >19.41    & PROMPT \\
2015-01-10 & 57032.34 &   >19.98    & PROMPT \\
2015-01-12 & 57034.26 &   >19.98    & PROMPT \\
2015-01-14 & 57036.31 & 20.30(0.32) & PROMPT \\
2015-01-16 & 57038.29 &   >20.02    & PROMPT \\ 
2015-01-21 & 57043.34 & 20.38(0.37) & PROMPT \\
2015-01-22 & 57044.26 &   >19.18    & PROMPT \\
2015-01-26 & 57048.32 &   >19.79    & PROMPT \\
2015-01-27 & 57049.26 & 20.19(0.36) & PROMPT \\
2015-01-29 & 57051.28 & 20.13(0.36) & PROMPT \\
2015-02-01 & 57054.26 &   >19.62    & PROMPT \\
2015-02-03 & 57056.33 &   >18.71    & PROMPT \\
2015-02-04 & 57057.27 &   >18.60    & PROMPT \\
2015-02-07 & 57060.15 &   >19.58    & PROMPT \\
2015-02-08 & 57061.30 &   >19.00    & PROMPT \\
2015-02-10 & 57063.10 &   >19.64    & PROMPT \\
2015-02-12 & 57065.22 & 20.36(0.25) & PROMPT \\
2015-02-17 & 57070.30 &   >19.75    & PROMPT \\
2015-02-18 & 57071.30 &   >19.83    & PROMPT \\
2015-02-22 & 57075.09 &   >19.78    & PROMPT \\
2015-02-24 & 57077.09 & 20.41(0.30) & PROMPT \\
2015-02-26 & 57079.27 &   >18.96    & PROMPT \\
2015-03-06 & 57087.08 &   >19.07    & PROMPT \\
2015-03-12 & 57093.14 &   >19.71    & PROMPT \\
2015-03-14 & 57095.18 & 20.49(0.33) & PROMPT \\
2015-03-18 & 57099.22 &   >19.08    & PROMPT \\
2015-03-28 & 57109.19 &   >19.06    & PROMPT \\
2015-03-30 & 57111.17 &   >19.02    & PROMPT \\
2015-04-04 & 57116.04 &   >19.15    & PROMPT \\
2015-04-11 & 57123.12 &   >19.14    & PROMPT \\ 
\hline
\end{tabular}
\end{table}

\end{document}